\documentclass{osa-article}
\pdfoutput=1
\journal{oe}

\articletype{Research Article}

\usepackage{algorithm}
\usepackage{algpseudocode}
\usepackage{svg}
\usepackage{lineno}
\linenumbers

\usepackage{listings}
\usepackage{xcolor}
\usepackage{rotating}
\usepackage{pdflscape}

\newcommand\scalemath[2]{\scalebox{#1}{\mbox{\ensuremath{\displaystyle #2}}}}
\allowdisplaybreaks

\usepackage{subfiles}

\begin{document}
\nolinenumbers

\title{Single pixel imaging at high pixel resolutions}

\author{Rafa{\l} Stojek,\authormark{1,2} Anna Pastuszczak,\authormark{1} Piotr Wr{\'o}bel,\authormark{1} and Rafa{\l} Koty{\'n}ski \authormark{1,*}}

\address{\authormark{1}University of Warsaw, Faculty of Physics, Pasteura 5, 02-093 Warsaw, Poland\\
\authormark{2}Vigo System, Poznańska 129/133, 05-850 Ożarów Mazowiecki, Poland\\}

\email{\authormark{*}Rafal.Kotynski@fuw.edu.pl} 



\begin{abstract}
The usually reported pixel resolution of single pixel imaging (SPI) varies between $32 \times 32$ and $256 \times 256$ pixels falling far below imaging standards with classical methods. Low resolution results from the trade-off between the acceptable compression ratio, the limited DMD modulation frequency, and reasonable reconstruction time, and has not improved significantly during the decade of intensive research on SPI. In this paper we show that image measurement at the full resolution of the DMD, which lasts only a fraction of a second, is possible for sparse images or in a situation when the field of view is limited but is a priori unknown. 
We propose the sampling and reconstruction strategies that enable us to reconstruct sparse images at the resolution of $1024 \times 768$ within the time of 0.3s. Non-sparse images are reconstructed with less details. The compression ratio is on the order of 0.4\% which corresponds to an acquisition frequency of 7Hz. Sampling is differential, binary, and non-adaptive, and includes information on multiple partitioning of the image which later allows us to determine the actual field of view. Reconstruction is based on the differential Fourier domain regularized inversion (D-FDRI). The proposed SPI framework is an alternative to both adaptive SPI, which is challenging to implement in real time, and to classical compressive sensing image recovery methods, which are very slow at high resolutions.
\end{abstract}

\section{Introduction}
Indirect image measurement techniques called single-pixel imaging (SPI) since their introduction over a decade ago~\cite{Duarte_2008,Cand_s_2006} have led to a considerable amount of novel ideas about image measurement at various wavelength ranges, spectral imaging, imaging through scattering media, 3D imaging etc.~\cite{Gibson2020,Edgar_2018}. Digital micromirror devices (DMD) are the most frequently used spatial light modulators in optical SPI set-ups and in this paper we will take into consideration their typical technical specifications such as resolution, modulation frequency, binary operation, and bandwidth. 
Modulation frequency of modern DMDs is on the order of twenty kilohertz. This is not a lot, as at least several thousand exposures are needed per single image measurement. A non-compressive sequential measurement at a resolution of $1024\times768$ would involve projecting patterns that take $77$GB at the bitrate of $17.7$Gb/s. As a result, 
 either the image acquisition time must be very long or the resolution is reduced to the commonly reported range between $32\times 32$ and $256\times 256$ falling far below imaging standards with classical methods. Additionally, for compressive imaging the time needed for digital image reconstruction may be substantial at higher resolutions.
 Most real-time reconstruction SPI approaches rely on a single-step image reconstruction using a fast transform, e.g. the Fourier (FFT), Walsh-Hadamard (FWHT), or Discrete-Cosine (DCT) Transforms, or on the evaluation of a matrix-vector product\cite{Czajkowski_2018,Stantchev2020}. There is also growing interest in using neural networks for image reconstruction or for removing artifacts caused by compression\cite{Higham2018,Rizvi2020,Wang:21}.

There exist alternatives to the DMD technology, which we will not consider in this paper. Far higher frame-rates are possible with structured illumination with LEDs arrays, although such setups for ghost imaging and SPI have been demonstrated only at low resolutions~\cite{Salvador-Balaguer2018,Xu2018,Wang_2020,Zhao_2019}. Modulation with rotary elements with fixed patterns is a high-speed cost-efficient alternative to using dynamic modulators in THz\cite{Guerboukha_2018,Chen2019,Hahamovich:2021}.
The modulation
speed may be also increased
by combining various light modulation techniques together or by using arrayed light sources
with a fast modulation rate\cite{Liu_oe_26_10048_2018,W.Yuwang2017}.
 High resolution images may be obtained by data fusion techniques when SPI is combined with high resolution images acquired with classical cameras~\cite{Soldevila:21,Ghezzi:21}. Block compressive imaging with use of multiple detectors or with a focal plane-array, as well as parallel detection with a pushframe camera also effectively increase the sampling frequency per pixel~\cite{Ke_2012,Mahalanobis_2014,Wu_2019,Stuart:IEEETCI-2021}.

To reach the resolutions above $128\times 128$ one has to accept strong compression (i.e.~a low compression ratio) which usually results in a poor image quality. One way of retaining a good image quality is to apply adaptive sampling~\cite{Phillips:17,Qian:19,Jiang:17,He:21} with a sequence of sampling patterns selected dynamically during the measurement. Having in mind the large bitrate at which subsequent patterns should be calculated and sent to the DMD, adaptive sampling is difficult to implement in real time and currently has a rather theoretical significance. 

In this paper we make an attempt to construct an SPI framework optimal in terms of using DMD modulation at high resolutions.
We will consider non-adaptive high resolution binary sampling and a low compression ratio. Similarly as in our recent work~\cite{Pastuszczak_2021} we will also assume that all sampling patterns contain approximately but not exactly half pixels in the on-state and that the measurement is differential. This provides an increased signal entropy and an improved signal-to-noise-ratio (SNR). We will also commence the image reconstruction with the differential Fourier domain (D-FDRI) regularization method proposed in~\cite{Pastuszczak_2021}. The novel elements proposed in this paper are the sampling patterns based on multiple image maps and the second stage of image reconstruction that makes use of these maps. Every map defines a distinct partitioning of all image pixels into nonoverlapping sectors.  In a noise-free scenario, the initial image reconstruction is guaranteed to provide the correct mean values for each sector of every map. The purpose of the second stage of image reconstruction is to determine the actual field of view based on the locations of empty sectors of every map. Non-empty sectors are corrected accordingly.
Overall, the initial image reconstruction provides a low-quality image which will be the final result for dense  images sampled at a low compression ratio. The second stage of the algorithm  improves the reconstructed image, potentially up to the resolution of the DMD, when the image is sparse.

\section{Sampling based on maps and an algorithm for image reconstruction and determination of the field of view}
\subsection{Objectives}
Our aim is to introduce a sampling and image reconstruction framework for SPI that would satisfy a number of practically driven requirements.

We want to make efficient use of the full DMD's spatial and frequency bandwidths for image sampling. For modern DMDs this implies binary sampling at the resolution on the order of $1024\times 768$ and at the modulation frequency of $22.5$~kHz. This corresponds to the transmission bandwidth of $17.7$~Gb/s. At such a high bandwidth, it is difficult to implement adaptive sampling. Therefore, our choice is to use binary non-adaptive sampling at a full DMD resolution. To make the measurement practically feasible, the acquisition time should not exceed a fraction of a second. This implies an extremely strong compression - for instance the compression ratio of $0.5\%$ corresponds to the acquisition rate of $5.6$~Hz, which is still reasonable. 

We want to use differential sampling. This is a simple technique to improve the SNR and to make the measurement independent of a constant or slowly varying background bias signal. At the same time, we do not want to compromise the available DMD bandwidth. We will use the differences between measurements with subsequent binary sampling patterns to reconstruct the image. This means that effectively we increase the number of measurements by one, rather than by a factor of $1/2$, $1/3$ or $1/p$ with $p\in \mathbb{N}$ that is often accepted with more straightforward differential methods~\cite{Radwell_2014,Zhang2015,Czajkowski_2019}.

We want to be able to determine the field of view composed of possibly non adjacent non-empty areas of the image easily.  With this aim, we propose a novel differential sampling scheme with binary sampling patterns that will serve a double role. First, the sampling should give the usual information about the spatial contents of the measured image. On top of this it should give some guarantees about the possibility of identifying empty regions of the image. We will consider multiple partitionings of the image surface with the help of auxiliary maps, and then construct the actual binary sampling patterns using these maps. By an image map we understand any partitioning of all image pixels into distinct subsets. These subsets may consist of both isolated as well as of neighbouring pixels. Sampling functions deduced from a single map should enable us to find the mean values of the measured image within every subset of the pixels in that map. Moreover, we want to keep approximately half of the pixels in the on-state in every sampling pattern.

Finally, we want to have a fast image reconstruction algorithm, capable of reconstructing the image at a quality dependent on image sparsity or on the field of view. When objects cover the whole surface of the image, the compression ratio is low (on the order of $0.5\%$ or less) and one can at most expect a low-quality reconstruction. When the field of view is limited, the quality may be improved.
The challenge is that the first situation requires low spatial frequency sampling, while the later, a high frequency sampling. In effect, neither random sampling with white noise, nor low-frequency e.g. Fourier or DCT sampling perform well. We propose to use image maps based on somehow arbitrarily chosen realizations of spatially correlated Gaussian noise to get a desired trade-off between low and high frequency sampling, and to construct the maps. 
 The reconstruction algorithm begins with a non-negative differential Fourier-domain regularized solutions (D-FDRI)~\cite{Pastuszczak_2021}. The advantages of the D-FDRI are that its implementation requires just one single matrix-vector multiplication, it is applicable with high resolution binary differential sampling, and it provides a regularized solution to the inverse problem. Following, the result is iteratively improved by assigning the value of zero to the determined areas of the image maps, and by scaling other pixels to repair the mean values within other regions. This procedure is iterated for all maps.

\subsection{Background on the differential Fourier Domain Regularized Inversion image reconstruction method (D-FDRI)\cite{Pastuszczak_2021,DFDRI:code}}
Compressive measurement $\mathbf{y}$ of an image $\mathbf{x}$ (with pixel values arranged in a vector) in the presence of additive signal noise $\mathbf{n_s}$ and detector noise $\mathbf{n_d}$ may be expressed as
\begin{equation}
	\mathbf{y}=\textbf{M}\cdot (\mathbf{x+n_s})+\mathbf{n_d},\label{eq.measurement}
\end{equation}
where the rows of the measurement matrix $\mathbf{M}$ contain the patterns displayed on the DMD during the measurement and $\cdot$ denotes the dot product. In~Eq.~(\ref{eq.measurement}) noise has been decomposed into a signal independent part $\mathbf{n_d}$ primarily attributed to the detector dark current and the signal dependent part $\mathbf{n_s}$ from sources such as background illumination\cite{Sun_2018}. The measurement is compressive when the dimension of $\mathbf{y}$ which will be denoted as $k$ is smaller than the number of pixels $n$ in the image $\mathbf{x}$. In our case, the sampling is binary and makes use of all the DMD pixels. Therefore, $\mathbf{M}$ is a binary matrix, $n=1024\cdot 768$, and $k$ is on the order of $3\cdot 10^3$. A linear reconstruction followed by the truncation of negative values (here denoted with a $ReLu$ function) takes the form
\begin{equation}
	\tilde{\mathbf{x}} = ReLu(\tilde{\mathbf{x}}_0),\text{ where }\tilde{\mathbf{x}}_0=\mathbf{P}\cdot \mathbf{y} .\label{eq.reconstr}
\end{equation}
Depending on the choice of the measurement matrix $\mathbf{M}$ and the reconstruction matrix $\mathbf{P}$, one can obtain various SPI schemes. A frequent choice is to take $\mathbf{M}$ and $\mathbf{P}$ as consisting of selected rows of a linear transform matrix and selected columns of its inverse eg. of a direct and inverse Fourier, DCT or Walsh-Hadamard transforms. In a more general approach, $\mathbf{P}$ may be the Moore-Penrose pseudoinverse $^+$ of $\mathbf{M}$ (i.e. $\mathbf{P}=\mathbf{M}^{+}$). The pseudoinverse approach allows for using an arbitrary form of sampling patterns but without regularization may result in a poor noise robustness. Even more generally, $\mathbf{P}$ may be any matrix that is a generalized inverse $^g$ of $\mathbf{M}$, (i.e. $\mathbf{P}=\mathbf{M}^{g}$). The generalized inverse is not unique and this approach allows for including some optimization or regularization in the choice of $\mathbf{P}$. A generalized inverse is used in the FDRI method~\cite{Czajkowski_2018} to regularize the inverse problem in the Fourier domain.
Finally, a regularization may be combined with a differential measurement and then $\mathbf{P}=(\mathbf{D}\cdot\mathbf{M})^{g}\cdot\mathbf{D}$, where $\mathbf{D}$ is the 1D finite-difference operator (discrete gradient $D_{i,j}=\delta_{i,j-1}-\delta_{i,j}$, where $\delta$ is the Kronecker delta).
This is what we have proposed in~\cite{Pastuszczak_2021} as the differential Fourier domain regularized inversion (D-FDRI) method, which is the starting point for the present work.

 We note that it is $\tilde{\mathbf{x}}_0$ and not $\tilde{\mathbf{x}}$ that fulfills the measurement equation (\ref{eq.measurement}) in the absence of noise. Still in practice, in terms of the peak signal to noise ratio (PSNR), the nonnegative image $\tilde{\mathbf{x}}$ is a better approximation to the object $\mathbf{x}$. 
For D-FDRI, the reconstruction matrix $\mathbf{P}$ is calculated from $\mathbf{M}$ as~\cite{Pastuszczak_2021},
\begin{equation}
	\mathbf{P}=\mathbf{F^*}\cdot\mathbf{\hat\Gamma}\cdot\mathbf{F}\cdot(\mathbf{D}\cdot\mathbf{M}\cdot\mathbf{F^*}\cdot\mathbf{\hat\Gamma}\cdot\mathbf{F})^{+}\cdot\mathbf{D},\label{eq.D-FDRI}
\end{equation}
where $\mathbf{F}$ is the 2D Fourier transform, $^*$ is the complex conjugate transpose, and the filter $\mathbf{\hat\Gamma}$ is a diagonal matrix~\cite{Czajkowski_2018}
\begin{equation}
	\hat\Gamma_{i,j}=\frac{\delta_{i,j}}{\sqrt{(1-\mu)^2(sin^2(\omega_x)+sin^2(\omega_y))+\mu^2\frac{\omega_x^2+\omega_y^2}{2\pi^2}+\epsilon }}\label{eq.Gamma},
\end{equation}
where $\mu$ and $\epsilon$ are used to tune the properties of the regularization and $\omega_{x,y}$ are the spatial frequencies. In the present paper we assume that $\mu=0.5$ and $\epsilon= 10^{-7}$. D-FDRI reconstruction may be combined with arbitrary measurement matrices, for instance with binary nonorthogonal patterns. D-FDRI is inherently differential, so the reconstruction is blind to the mean value of the noise $\mathbf{n_d}$. It could be also directly applied with complementary sampling schemes if the SNR needs to be further improved\cite{Pastuszczak_2021}. Calculation of the reconstruction matrix $\mathbf{P}$ is computationally and memory intense but is done once. Afterwards, the image is reconstructed from the measurement $\mathbf{y}$ with use of Eq.~(\ref{eq.reconstr}) in single matrix-vector multiplication.

\begin{figure}[!ht]
	\centering\includegraphics[width=14cm]{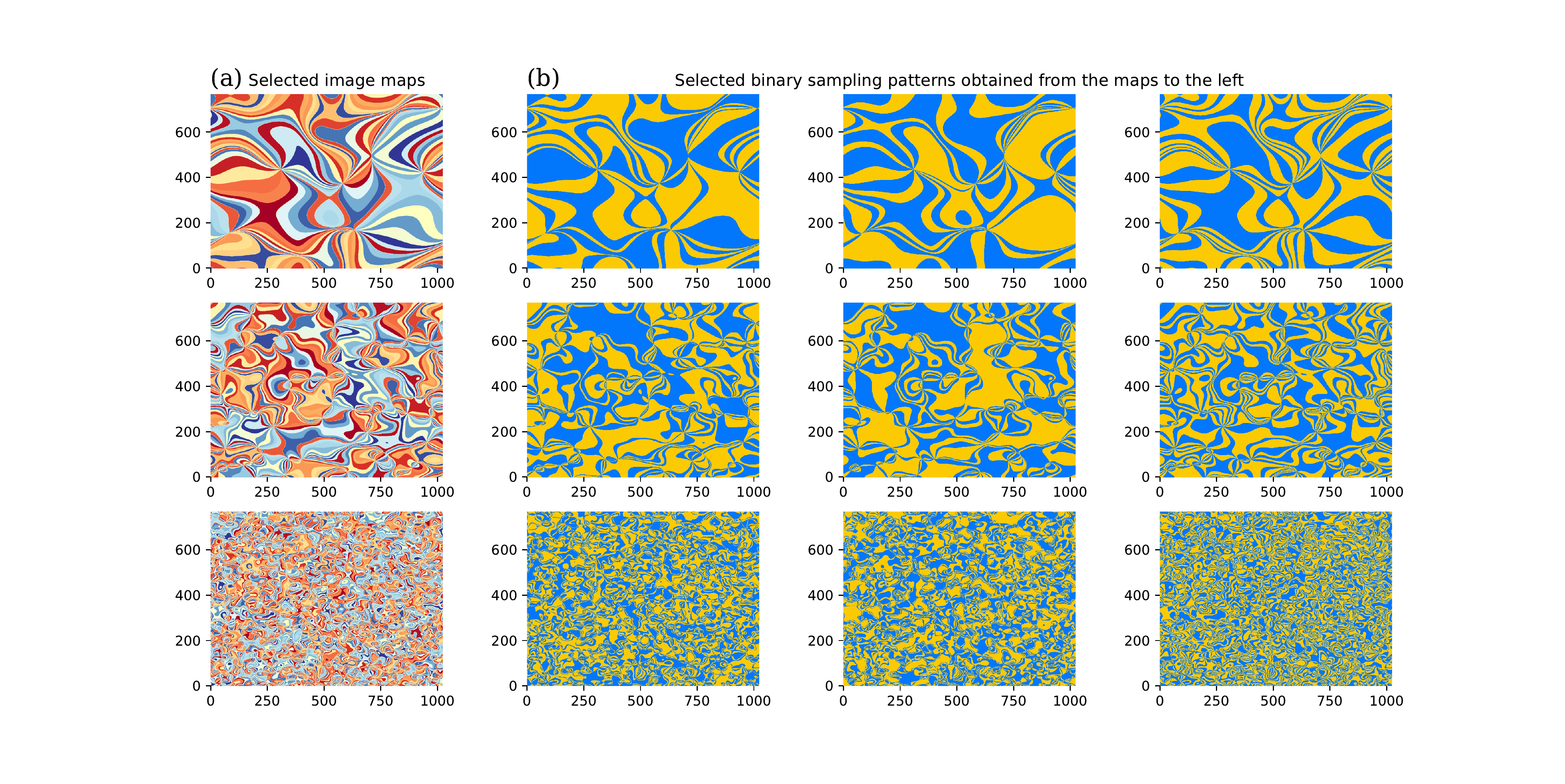}
	\caption{Examples of high resolution binary sampling patterns and of auxiliary maps from which they have been derived. (a)~Selected maps obtained by uniform quantization of the phase of correlated complex Gaussian noise into $m=31$ discrete levels. The three maps shown are selected from a set of $l=100$ maps, each of which was obtained from a Gaussian noise with a distinct randomly chosen autocorrelation length. (b)~Selected  $1024\times 768$ binary sampling patterns obtained with the help of the maps shown in the corresponding row of subplot (a).}\label{fig.maps1}
\end{figure}

\begin{figure}[!ht]
	\centering\includegraphics[width=14cm]{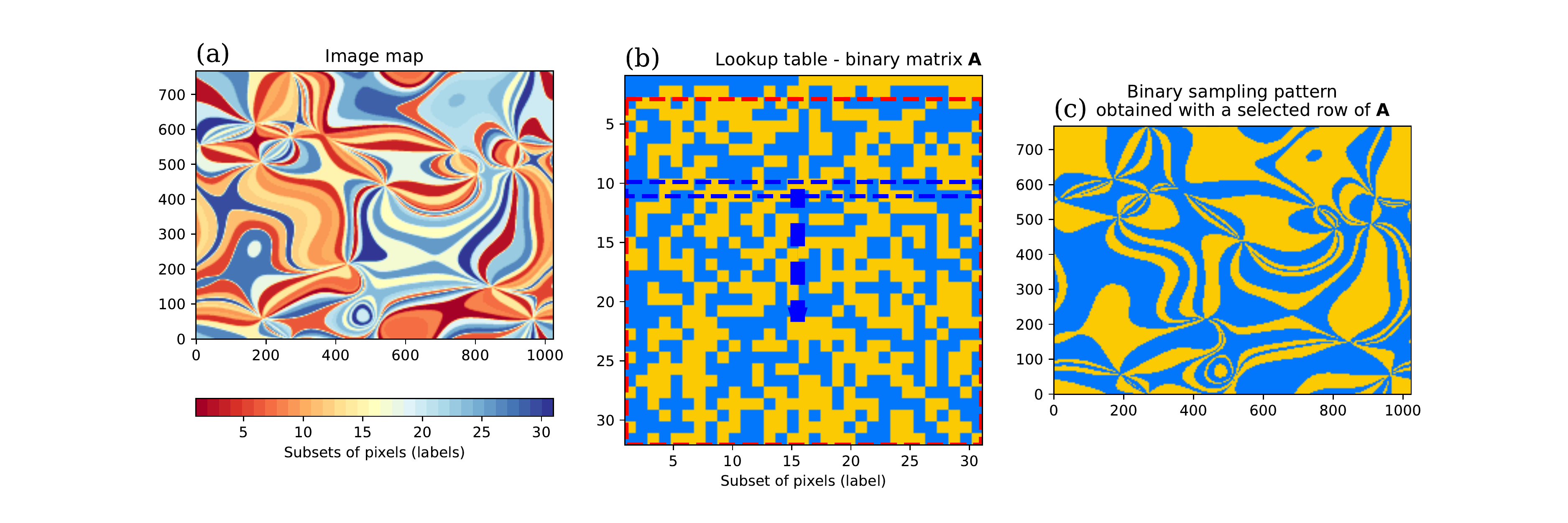}
	\caption{This figure explains how image maps are converted into binary sampling patterns with the help of the auxiliary lookup table $\mathbf{A}$. (a)~A discrete $m$-level image map taken from the sequence of $l$ maps (here $m=31$, and $l=100$);  (b)~The auxiliary binary matrix $\mathbf{A}$ whose $m+1$ rows are used as a sequence of lookup tables for conversion of every map into up to $m+1$ binary patterns  (blue and yellow denote the values of $0$ and $1$; see the Supplementary Materials for a list of lookup tables with different sizes). (c)~The binary sampling pattern obtained from the map in subplot (a) using the row of matrix $\mathbf{A}$ marked with a blue dashed rectangle. Pattern encoding is repeated for subsequent rows of $\mathbf{A}$ and for subsequent maps. The first two rows of $\mathbf{A}$ are used only for the first map and are omitted for other maps to remove linear dependence from the set of sampling patterns. Overall this gives a sequence of $k=(m+1)+(l-1)\cdot(m-1)=3002$ binary differential $1024\times768$ sampling patterns that form the rows of the measurement matrix $\mathbf{M}$.}\label{fig.maps2}
\end{figure}

\subsection{Image maps}
Our measurement matrix $\mathbf{M}$ consisting of rows with binary sampling patterns is calculated based on a set of image maps $\mathbf{m}_1, \mathbf{m}_2,... \mathbf{m}_{l}$. Every map consists of the same number of pixels as the DMD and is defined by labeling the pixels with integer numbers between $1$ and $m$, i.e. $\mathbf{m}_i\in \lbrace 1,...m\rbrace^n$, with $i\in\lbrace 1,l \rbrace$ . In this paper the number of maps is $l=100$, and $m=31$ but other odd values of $m$ are also possible. Our maps are arbitrarily defined with the only assumption that they should be dominated with low spatial frequency contents but still contain some high spatial frequency information as well. A more extensive study of possible map compositions is beyond the scope of this paper but certainly some further optimization of the maps for a given kind of images is possible. Here the maps are obtained by generating spatially correlated Gaussian complex-valued noise and by assigning the label $j\in\lbrace 1,m \rbrace$ to each pixel based on the uniform quantization of the phase level of a noise realization. 
Some samples of the maps are illustrated in Fig.\ref{fig.maps1}(a). The resolution of these maps is $1024\times 768$ and each of them defines an image partitioning into $m=31$ pixel subsets. Then, each map is translated into a sequence of binary patterns which are stacked together to form the measurement matrix $\mathbf{M}$. Fig.~\ref{fig.maps1}(b) illustrates some of the patterns created using the map from Fig.~\ref{fig.maps1}(a).

Translation of maps into binary patterns is graphically explained in Fig.~\ref{fig.maps2}. Translation is performed with the help of subsequent rows of an auxiliary binary matrix $\mathbf{A}$ which play the role of lookup tables. To create a single sampling pattern we select one row from $\mathbf{A}$ and replace the $j$-th region in the map (with $j\in\lbrace 1,..m\rbrace$) with the binary value taken from the $j$-th column of the lookup table. Then we repeat the same using subsequent rows from $\mathbf{A}$. 
 Matrix $\mathbf{A}$ is shown in Fig.~\ref{fig.maps2}(b) in a graphical form with the values of $0$ and $1$ indicated with yellow and blue . $\mathbf{A}$ contains $m+1$ rows and $m$ columns so a sequence of $m+1$ sampling patterns could be obtained from a single map.
 However, only the first map is used to create $m+1$ binary patterns, and each of the following maps is translated into $m-1$ patterns (by omitting the first two rows of $\mathbf{A}$). This assures that the differential measurement includes nonredundant information about the mean intensity of every region of every map and that the sampling patterns are linearly independent. Otherwise, for instance the information about the mean value of the entire image could be found independently using the sampling patterns obtained from every map.  Overall, the number of binary sampling patterns equals $k=(m+1)+(l-1)\cdot(m-1)$, the number of measurements effectively used for image reconstruction equals $k-1$ (one measurement is lost due to the differential treatment of data), and the number of sectors in all the maps is equal to $l\cdot m$ (which exceeds $k$).
 
 Matrix $\mathbf{A}$ is created in the same way as in \cite{Pastuszczak_2021} but is used differently. $\mathbf{A}$ is obtained by brute force numerical search under the condition 
 that rank of the matrix obtained by subtracting subsequent rows of $\mathbf{A}$ is equal to $m$, i.e. $rank(\mathbf{D}\cdot\mathbf{A})=m$, and that each of the rows of $\mathbf{A}$ contains $m\pm 1$ ones and zeros. 
 As a result, the proposed differential sampling probes independently every region of the map, it is binary, and it consists of patterns with approximately half of the pixels in the on and off states. These properties of sampling patterns are advantageous in terms of signal entropy and sensitivity to truncation noise~\cite{Pastuszczak_2021}. 
 
 Let us introduce a denotation for the set of indexes pointing to image pixels that belong to $j$-th sector of $i$-th map
 \begin{equation}
 	{\mathbf{v}}_{(i,j)}=\lbrace p : [\mathbf{m}_i]_p=j\rbrace,\text{ where } i\in\lbrace 1..l\rbrace, j\in\lbrace 1..m\rbrace
 \end{equation}
where $p\in\lbrace 1,..n\rbrace$ enumerates the pixels.

 A useful property of the proposed sampling is that in a noise-free scenario an image $\tilde{\mathbf{x}}_0$ reconstructed from the compressive differential measurement using formula ~(\ref{eq.reconstr}) will have the correct mean values calculated within every region of every map, i.e.
 \begin{equation}
 	\overline{\tilde{\mathbf{x}}_0 [\mathbf{v}_{(i,j)}]} =	\overline{{\mathbf{x}} [\mathbf{v}_{(i,j)}]}.
    \label{eq.map_mean_values}
 \end{equation} 
This is true because $\tilde{\mathbf{x}}_0$ satisfies the noise free measurement equation, and the differential sampling with proposed patterns based on maps is equivalent to non-differential sampling with patterns consisting of all individual sectors extracted from the maps.

 For nonnegative images $\mathbf{x}$, finding regions with mean value of $\tilde{\mathbf{x}}$ (approximately) equal to zero enables us to mark them as empty and to eliminate them from image reconstruction.

\subsection{Reconstruction algorithm}
\label{sec.rekonstr}
The proposed algorithm consists of two stages. First, an initial approximate reconstruction $\tilde{\mathbf{x}}$ is calculated using equation (\ref{eq.reconstr}) with the reconstruction matrix $\mathbf{P}$ given by Eqs.~(\ref{eq.D-FDRI}) and (\ref{eq.Gamma}). For dense images this will be also the final result. In the second stage, empty image regions are identified from the mean values calculated over every sector of every map for which $\overline{\tilde{\mathbf{x}}_0 [\mathbf{v}_{(i,j)}]}\le\epsilon_n$ and pixels from these sectors are set to zero. If this change affects some sector of another map, the remaining pixels from that sector are scaled to restore the proper mean value. The algorithm is detailed below.

\begin{algorithm}
\caption{MD-FDRI image reconstruction algorithm}\label{alg:md-fdri}
	\begin{algorithmic}[1]
		\Function{MD-FDRI}{$\mathbf{y},\mathbf{P}, \mathbf{v}, p , n$}\Comment{Image reconstruction algorithm}
		\State $\tilde{\mathbf{x}}_0 \gets \mathbf{P}\cdot\mathbf{y}$ 
	 \State $\tilde{\mathbf{x}} \gets ReLu(\tilde{\mathbf{x}}_0)$\Comment{Initial D-FDRI reconstruction}
		\For{$i=1\text{ to } l$} \Comment{Loop over maps}
		\For{$j=1\text{ to } p$} \Comment{Parallel loop over sectors of map $\mathbf{m_i}$}
		\If {$\overline{\tilde{\mathbf{x}}_0 [\mathbf{v}_{(i,j)}]}<\epsilon_{n}$}	\Comment{Is the sector empty? ($0<\epsilon_{n}<<1$) }				
		\State $\tilde{\mathbf{x}} [\mathbf{v}_{(i,j)}]\gets\mathbf{0}$ \Comment{If yes, set it to zero}
		\Else
		\State $\tilde{\mathbf{x}} [\mathbf{v}_{(i,j)}]\gets\tilde{\mathbf{x}} [\mathbf{v}_{(i,j)}] \frac{\overline{\tilde{\mathbf{x}}_0 [\mathbf{v}_{(i,j)}]}}{\overline{\tilde{\mathbf{x}} [\mathbf{v}_{(i,j)}]}}$ \Comment{Otherwise, correct the sector's mean}
		\EndIf
		\EndFor
		\EndFor
		\State \textbf{return} $\tilde{\mathbf{x}}$ \Comment{Return the reconstructed image}
		\EndFunction
	\end{algorithmic}
\end{algorithm}
 The numerical cost of this algorithm $O(n\cdot k)$ is proportional to the number of pixels $n$, so the proposed method may be used with high resolution images. At the same time, the compression ratio $k/n$ has to be low, otherwise the matrix $\mathbf{P}$ with $n\cdot k$ elements becomes too large to be calculated or even stored in computer memory. Later in this paper we will use the denotation MD-FDRI for the proposed modified map-based D-FDRI method.
 
\subsection{Discussion}
 MD-FDRI gives a high quality reconstruction when the image is spatially sparse. For gray-scale dense images it works reasonably well but other SPI methods may give better results. The   optimal choice of the parameters $l$ an $m$ and of the geometric shapes of map regions is object dependent and is beyond the scope of this paper.  It  should somehow reflect the complexity of the geometric shapes of the sparse regions of images. The most desired situation is when as many region maps as possible entirely overlap with sparse parts of the measured image.   In fact, a comparison of results obtained with different values of $l$ and $m$ in the Sect. S5 of the Supplementary Materials indicates that the reconstruction quality does not vary in a simple or regular way with $l$ and $m$.  $m=31$ was the largest value for which we were able to calculate the lookup table $\mathbf{A}$ using brute force optimization and this value works well for various kinds of sparse images. 
 
 It may be interesting to make it clear why the lookup table  $\mathbf{A}$ is at all needed to encode the sampling functions instead of directly using separate image regions as distinct sampling functions. Numerically, the two approaches would give similar results. However, with the MD-FDRI lookup-table encoding,  the optical signal measured with a photodiode is higher by a factor of approximately $m$/2=31/2=15 which is substantial. This is because approximately half of the DMD pixels take the “on” or “off” positions for each sampling pattern, while the regions of image maps contain only $1/31$ of pixels.
A similar result could be obtained with a Hadamard matrix used as the lookup table.  However unlike the proposed lookup table, a Hadamard matrix contains a single row consisting of ones, and other rows with half of the elements containing ones. This leads to approximately twice larger signal for one sampling pattern than for the others, which unnecessarily raises the demand for the bit-resolution of the A/D converter used in the measurement. Our lookup table only consists of rows with approximately the same numbers of ones and of zeros.
On top of this,  using the lookup table $\mathbf{A}$  assures that information about the map regions is  found from a differential measurement, and if the detector adds systematically some constant bias to the detection signal, this bias does not affect the reconstruction result, while at the same time, the mean value of the image is still measured correctly. 
 
 In this paper we pay special attention to the kind of sparsity obtained by limiting the field of view of the observed image. This is practically interesting because by using a diaphragm in an optical set-up one could improve the image quality within the limited FOV without any other modifications to the set-up, sampling functions or to the reconstruction algorithm which would still work at the original resolution ($n=1024\times 768$). A full resolution dense image can not be measured accurately at the compression ratio of $0.4\%$ but one could scan the image with a diaphragm in a sequence of measurements to obtain a high quality measurement.
 
More detailed information on the generation of image maps, on constructing the measurement matrix from the image maps, on the numerical form of the lookup tables with different sizes and on removing linear dependence from the sampling patterns can be found in the Supplementary Materials.

\section{Numerical results}
Later in this paper we will use the denotation MD-FDRI for the modified map-based D-FDRI method.
For the numerical and experimental validation of the proposed MD-FDRI method, we use sampling patterns at the full resolution of the DMD present in our setup, i.e. $n=1024\times 768$. 
The parameters of the image maps used for generating the sampling functions are selected as $l=100$ and $m=31$ (some examples using different sets of parameters $l$ and $m$ are also presented for comparison in our Supplementary Materials). 
The resulting number of the sampling patterns equals $k=3002$ and the compression ratio is on the order of $CR = k/n \approx 0.4\%$. The image maps are generated by quantization (with $m$ discrete value levels) of the phase of spatially correlated complex Gaussian noise with the standard deviation of the Gaussian functions selected separately for each map according to the empirical formula $\sigma_j = 0.01+0.09 {r_j}^2$, where $j=1,2,...,m$ is the index of the map and $r_j$ are selected randomly from a uniform distribution on range $[0,1]$.

\begin{figure}[th!]
\centering\includegraphics[width=\linewidth]{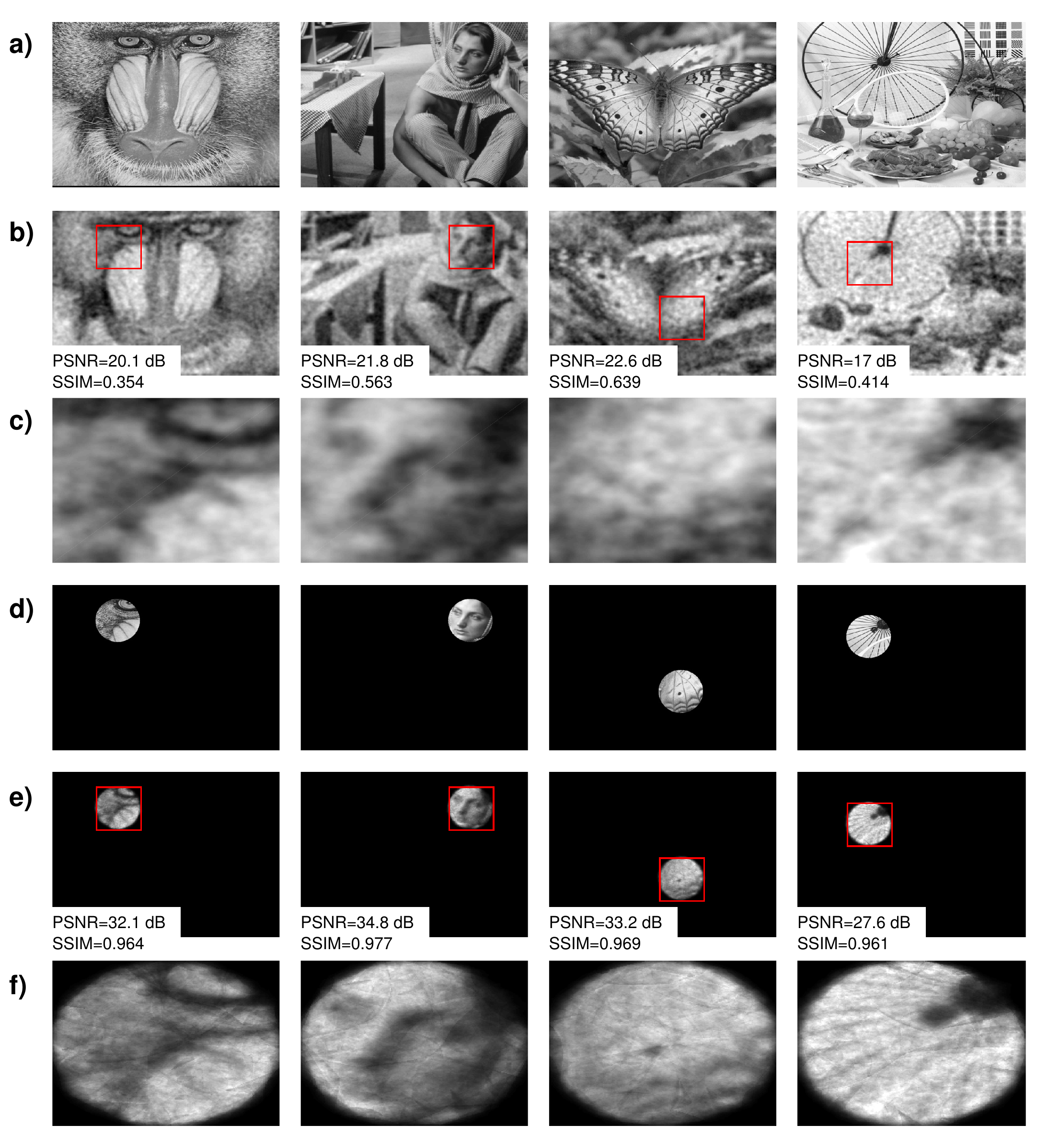}
\caption{Exemplary MD-FDRI reconstructions of several test images with resolution $1024 \times 768$ and either unlimited or reduced field of view. (a) Ground truth images. (b) MD-FDRI reconstructions for simulated SPI measurement of images (a). Fragments marked in red are enlarged in (c). (d) Sparse images obtained by limiting the field of view ($\gamma = 0.04$) of images in subplot (a). (e) MD-FDRI reconstructions for simulated SPI measurement of images (d). Fragments marked in red are enlarged in (f).}
\label{fig:rekostr_num}
\end{figure}

\begin{figure}[th!]
\centering\includegraphics[width=14cm]{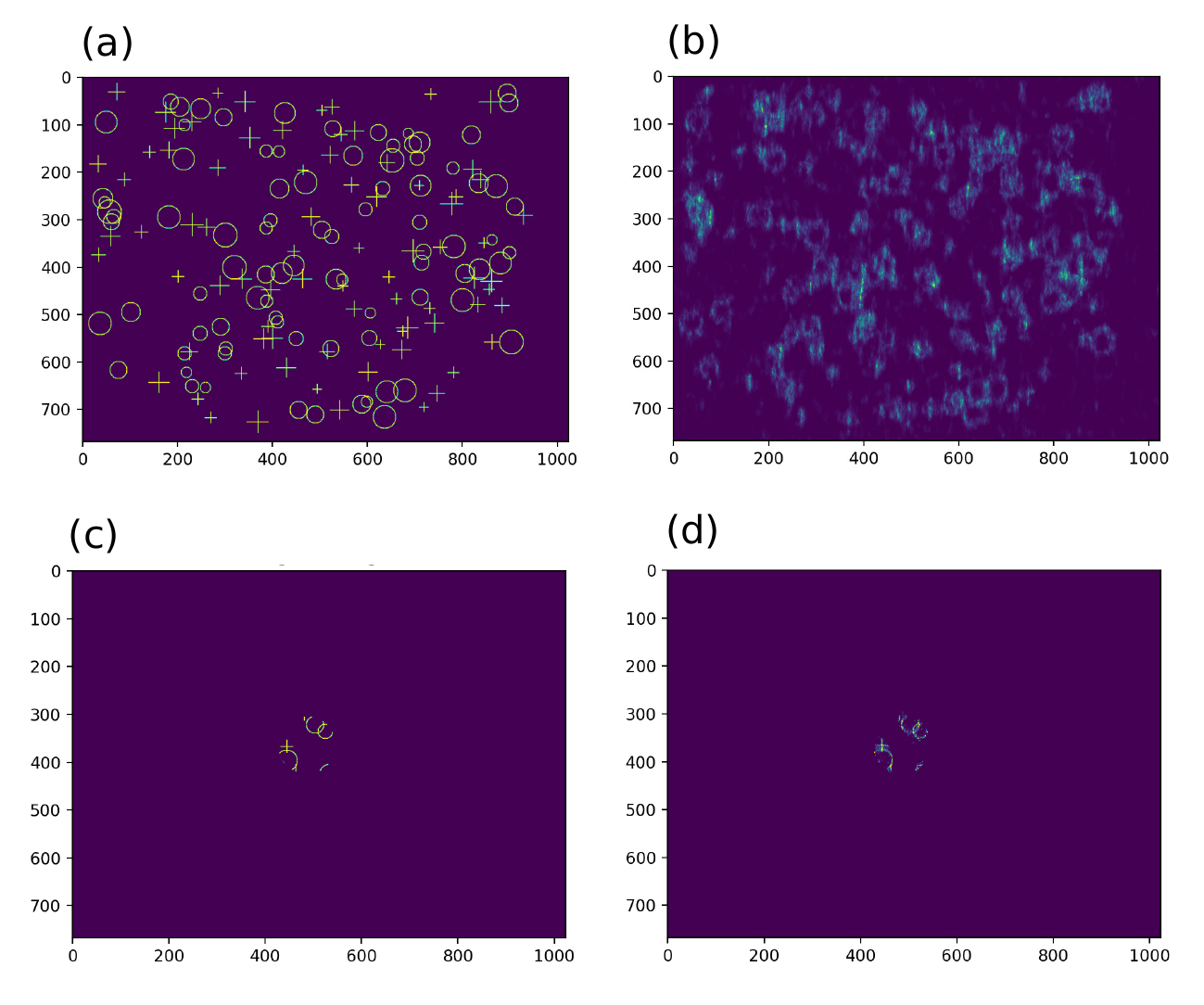}
\caption{SPI imaging of a high resolution image measured through a simulated diaphragm with a varied aperture. A complete simulation is included in the Visualisation 1 and selected frames are shown in subplots (b) and (d). (a),(c)~the ground truth (without and with the diaphragm). (b),(d)~corresponding images reconstructed with the proposed algorithm from a compressive measurement at the compression ratio of $k/n=0.4\%$. These two images as well as every frame in Visualisation 1 are reconstructed each from a separate differential compressive measurement with exactly the same $k=3002$ binary sampling patterns. For the modulation frequency of $22.5$kHz, the measurement time is $0.14$s. The reconstruction time of a single frame is on the order of $0.3$s. Sampling is nonadaptive and the algorithm does not need to know the location of the diaphragm. This example shows a realistic way to obtain SPI at the full resolution of the DMD with image measurement and reconstruction times on the order of a fraction of a second.}\label{fig:ScreenZFilmu}
\end{figure}

In Fig.~\ref{fig:rekostr_num} we present exemplary reconstructions of several test images obtained with the proposed MD-FDRI method in a simulated SPI measurement. The complete image set used in the simulation contains 50 non-sparse test images. We further consider both sparse and non-sparse images. The sparse images are obtained from the same image set by limiting the visible field of view, i.e. by replacing values of all pixels in the image, apart from a selected image area, with zeroes.
To compare images with different field of view, we introduce a parameter $\gamma$ ($0 \leq \gamma \leq 1$), which denotes the proportion of the image area which contains relevant visual information. The placement of the non-empty regions in the image may be arbitrary. 
We note that to some extent, the significance of parameter $\gamma$ is similar to a measure of image sparsity, however they are not strictly equivalent, as the non-empty parts of the image may still be compressible.
The reconstructions of both sparse and non-sparse images obtained with the MD-FDRI method show good quality, considering the extreme compression ratio used in the SPI sampling. Moreover, by reducing the field of view to a small region of the image, the quality of reconstruction of this region is strongly enhanced, as shown in Fig.~\ref{fig:rekostr_num}(d-f).

Another example of the performance of MD-FDRI is shown in Visualization 1 and selected frames from the visualization are presented in Fig.~\ref{fig:ScreenZFilmu}. A synthetic high resolution image filled with a large number of objects is measured through a simulated diaphragm with a varied aperture. A similar diaphragm could find application in SPI microscopy, where the aperture size and location could be moved to the locations of interest which will be then measured at the full resolution of the DMD. The trade off between resolution and field of view is something common to various imaging techniques involving an information channel with a limited bandwidth~\cite{Lukosz:JOSA-1967-57-932,Sheppard:Micron-2007-38-165}, and a similar trade-off is known in classical microscopy as well. We see that although the same non-adaptive sampling patterns are used for all frames, the resolution of reconstructed images vary, and approaches the resolution of the DMD for small aperture sizes.

\begin{figure}[th!]
\centering\includegraphics[width=\linewidth]{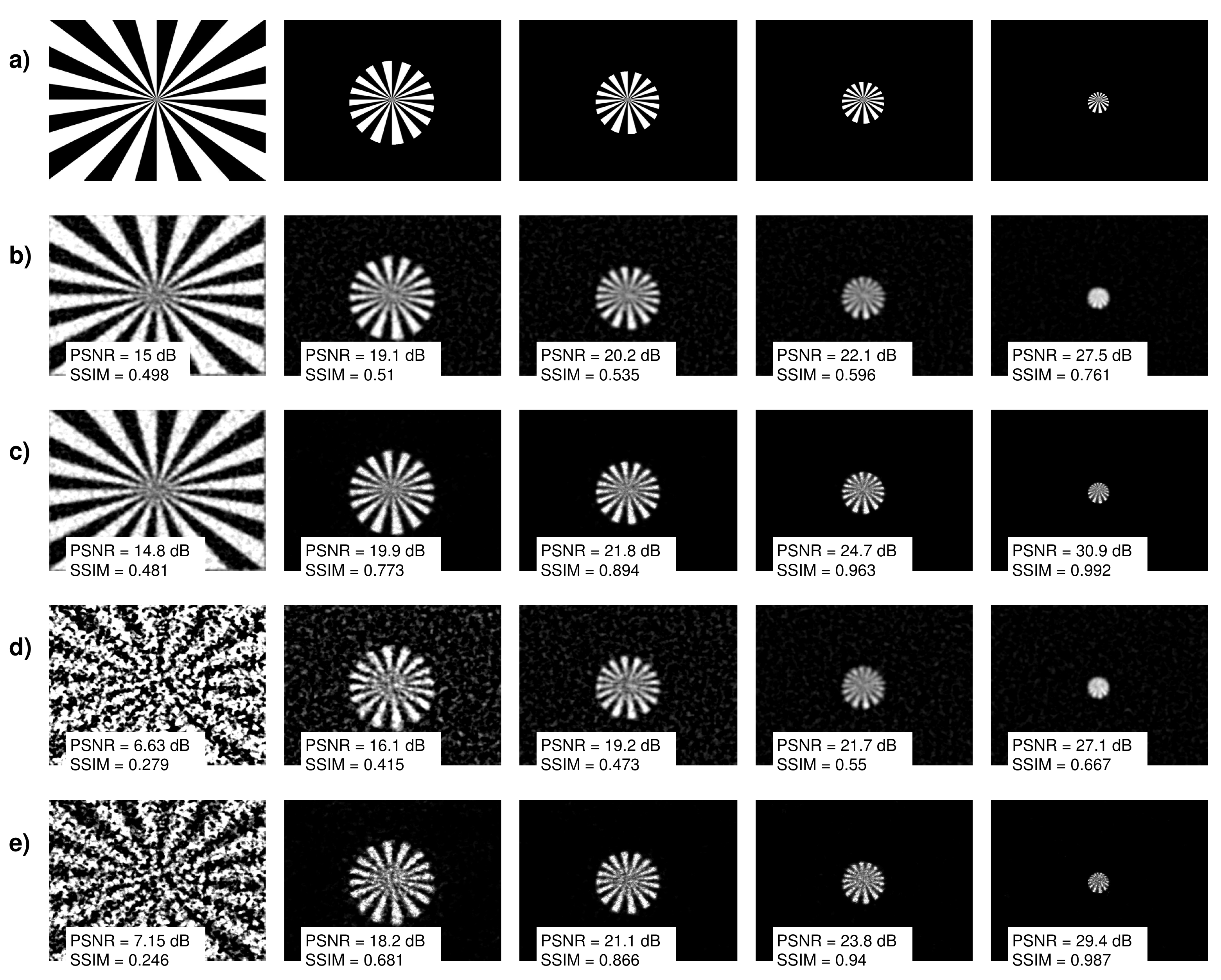}
\caption{Comparison of the reconstructions obtained with the first stage only and with both stages of the proposed MD-FDRI reconstruction algorithm for simulated SPI measurements. (a) Five ground truth images of a Siemens star resolution chart with different sizes of the field of view: $\gamma=1$, $\gamma=0.16$, $\gamma=0.09$, $\gamma=0.04$, and $\gamma=0.01$, respectively. (b-c) First stage (b) and final (c) reconstruction in a noiseless measurement scenario. (d-e) First stage (d) and final (e) reconstruction for a measurement in the presence of noise. An additive white Gaussian noise model is used with the relative standard deviation $\sigma=5\cdot 10^{-4}$. }
\label{fig:siemens-star}
\end{figure}

\begin{figure}[th!]
\centering\includegraphics[width=6cm]{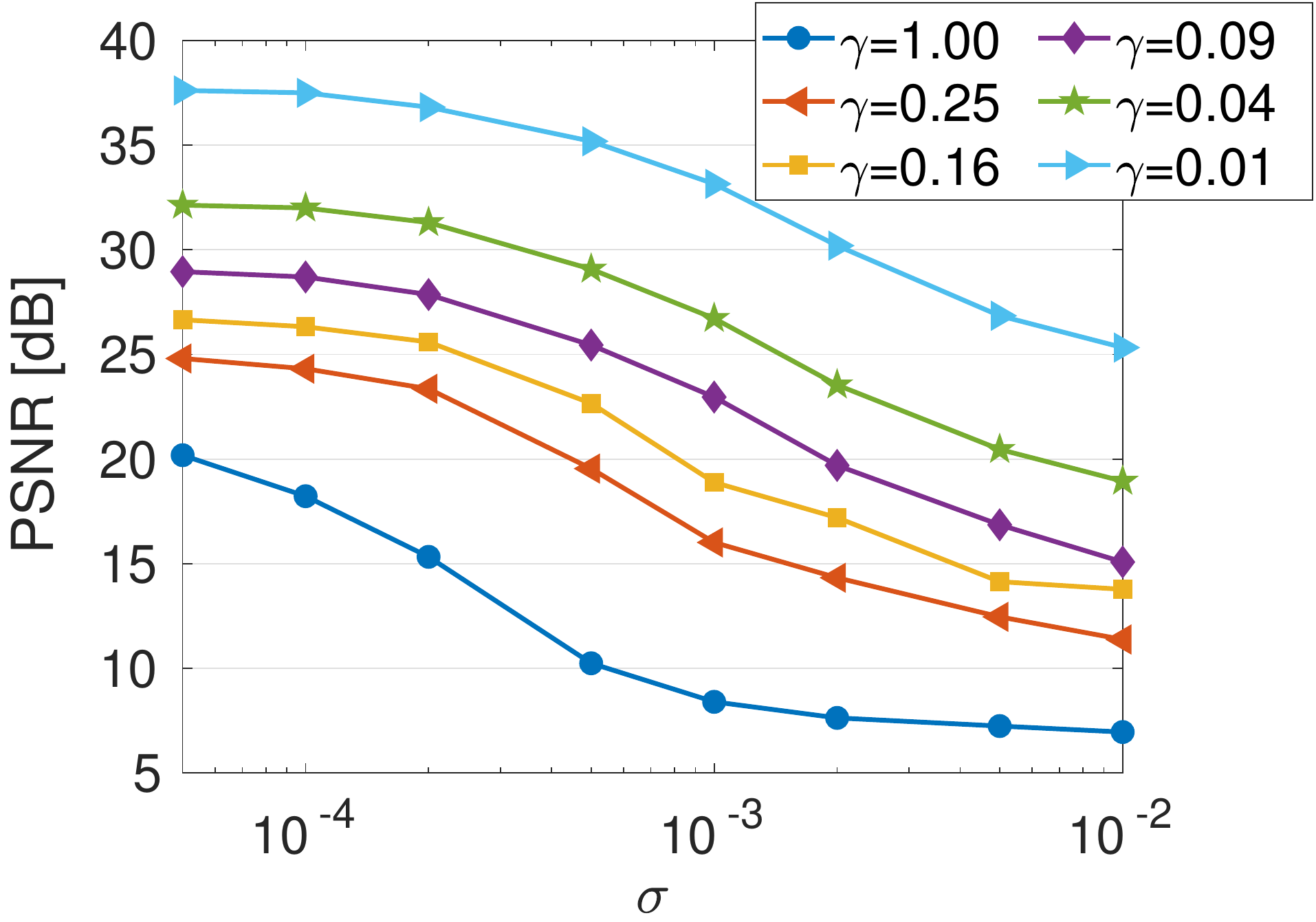}
\caption{Noise robustness of MD-FDRI for simulated measurements with additive white Gaussian noise with relative standard deviation $\sigma$. Average PSNR calculated over 50 test images for each size of the field of view $\gamma$.}
\label{fig:PSNR_noise}
\end{figure}

The enhancement of the reconstruction quality for sparse images is obtained in the second stage of the MD-FDRI reconstruction. MD-FDRI consists of two stages. The first stage produces an initial reconstruction with a single matrix-vector product using Eq.~(\ref{eq.reconstr}), similarly to our previous work \cite{Pastuszczak_2021}. In the second stage, the algorithm iteratively clears empty sectors of the image and corrects the mean value in each non-empty region of every image map. Therefore, the second stage of MD-FDRI yields the better reconstruction quality, the more empty sectors the image actually contains. 
Fig.~\ref{fig:siemens-star} presents the comparison of the first stage and final MD-FDRI reconstructions obtained for images with different values of $\gamma$. In this example, a standard Siemens star resolution chart is used as a test object. With narrowing of the field of view of the image, the effective resolution of the final MD-FDRI reconstruction improves, while for the first-stage reconstruction it remains unchanged.
Fig.~\ref{fig:siemens-star}(d-e) also shows the influence of additive measurement noise on the reconstruction quality obtained at each stage. Presence of noise interferes with the measurement of the correct mean values of each of the image regions, into which the images are partitioned by the maps. Therefore, Eq.~(\ref{eq.map_mean_values}) no longer holds true. The second stage of the reconstruction algorithm still allows us to remove the reconstruction artifacts from the empty sectors of the image. However, some improvement to the reconstruction quality of the non-empty regions is possible only if the SNR of the measurement is relatively high. 
The robustness of the MD-FDRI method to additive measurement noise is further illustrated in Fig.~\ref{fig:PSNR_noise}.

\begin{figure}[th]
\centering\includegraphics[width=\linewidth]{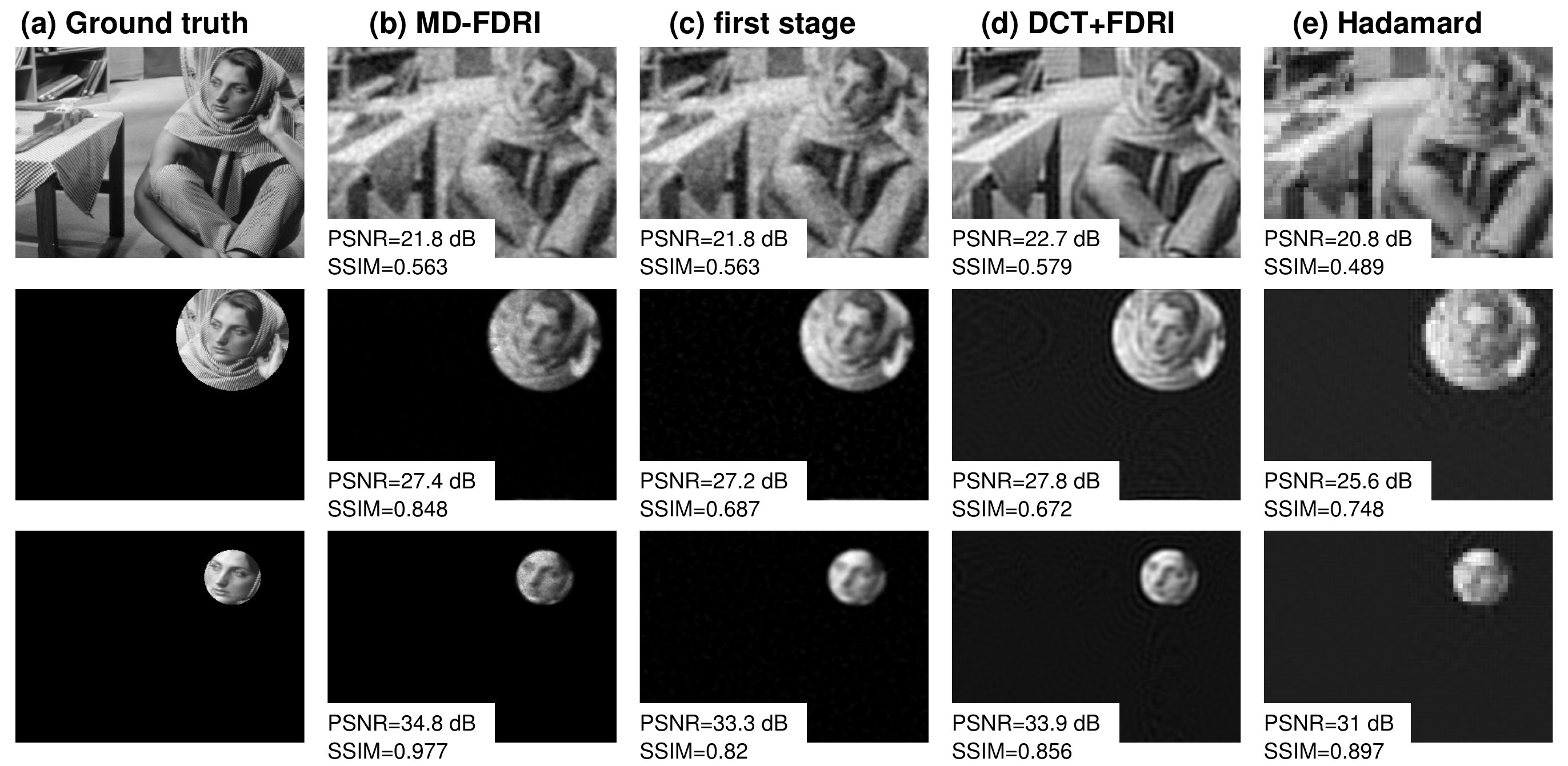}
\caption{Comparison of the proposed MD-FDRI reconstruction method to other sampling and reconstruction scenarios with the same image resolution $1024\times 768$ and compression ratio $CR=0.4\%$. (a) Ground truth images with different sparsity ($\gamma=1, \gamma=0.16, \gamma=0.04$ respectively). (b) Final MD-FDRI reconstructions. (c) First stage MD-FDRI reconstructions. (d) Sampling with binarized DCT elements and FDRI \cite{Czajkowski_2018} reconstructions. (e) Sampling with Walsh-Hadamard patterns and reconstruction with inverse transform. For DCT and Walsh-Hadamard sampling, low frequency elements of the respective transforms are selected.}
\label{fig:rec_compare}
\end{figure}

\begin{figure}[th]
	\centering\includegraphics[width=12cm]{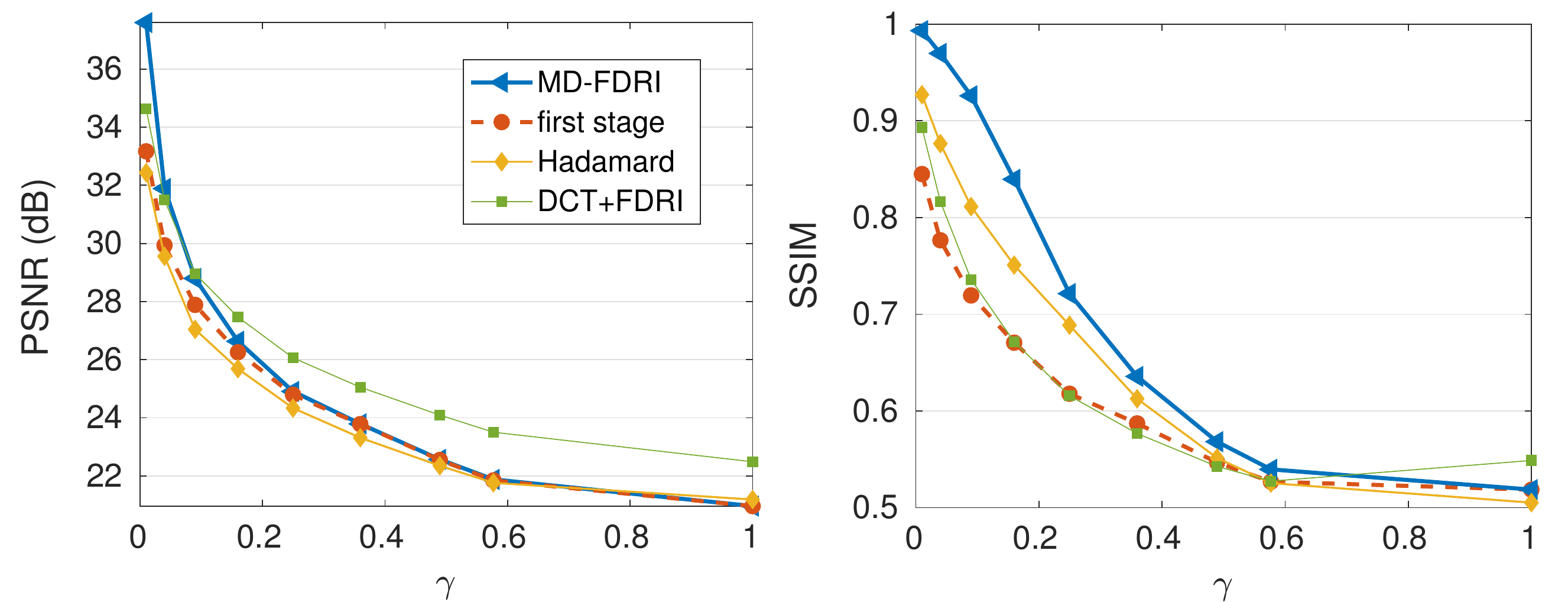}\caption{Average PSNR and SSIM of the reconstructions obtained with MD-FDRI as compared to other SPI sampling and reconstruction scenarios with the same image resolution and compression ratio. Simulated SPI measurement of 50 test images with gradually reduced field of view $\gamma$.}
	\label{fig:psnr_gamma}
\end{figure}

Finally, in Fig.~\ref{fig:rec_compare} and Fig.~\ref{fig:psnr_gamma} we compare the proposed map-based SPI sampling and MD-FDRI reconstruction routine to two other commonly used sampling and reconstruction scenarios. The first one comprises Walsh-Hadamard (WH) sampling patterns and the inverse transform used for image reconstruction. In the second scenario, binarized DCT sampling patterns are used instead and the reconstruction is obtained using FDRI method, similarly to our previous work \cite{Czajkowski_2018}. In each case, the sampling matrix consists of patterns at full resolution of the DMD and the compression ratio is $0.4\%$. 
The reconstructed images are compared in terms of two metrics: the peak signal-to-noise ratio (PSNR) and the structural similarity index (SSIM). 
While DCT+FDRI seems to offer the best reconstruction quality for SPI imaging of non-sparse images, in the case of images with high sparsity, MD-FDRI produces much richer reconstructions with significantly more detail than the other considered methods. For $\gamma=0.01$, average PSNR obtained with MD-FDRI is by over $3~dB$ higher than for DCT+FDRI and by over $5~dB$ higher than for WH. In general, the reconstructions obtained with MD-FDRI are usually of higher quality that the ones obtained with WH for most values of $\gamma$.
We note,  that the two metrics: PSNR and SSIM are not always consistent in evaluation of the reconstructions, especially of sparse images. For instance, for some values of $\gamma$ the reconstructions obtained with DCT+FDRI have the highest PSNR of all considered methods and the lowest SSIM at the same time. Also MD-FDRI outperforms other methods for much broader range of $\gamma$ in terms of SSIM than in terms of PSNR. This is because, these two metrics differently weight the reconstruction errors occurring in the empty regions of the image. The empty sectors are on average reconstructed with much smaller absolute value of the errors than the non-empty ones, but the relative error is still large. According to the PSNR criterion, the reconstruction errors are averaged over the whole image, therefore mostly the accuracy of reconstructing the non-empty sectors of the image contributes to this metric. On the other hand, for SSIM criterion the errors are averaged locally within a Gaussian window around each pixel of the image and the final metric is the mean value of the local ones. Therefore, the accuracy of reconstructing both empty and non-empty areas of the image are equally important, when this criterion is used. The advantage of MD-FDRI over other reconstruction methods in terms of SSIM visible in Fig.~\ref{fig:psnr_gamma} reflects the unique ability of this method to locate the empty sectors in the recovered images and clear them from reconstruction artifacts.

\section{Optical results}
\begin{figure}[h!]
\centering\includegraphics[width=10cm]{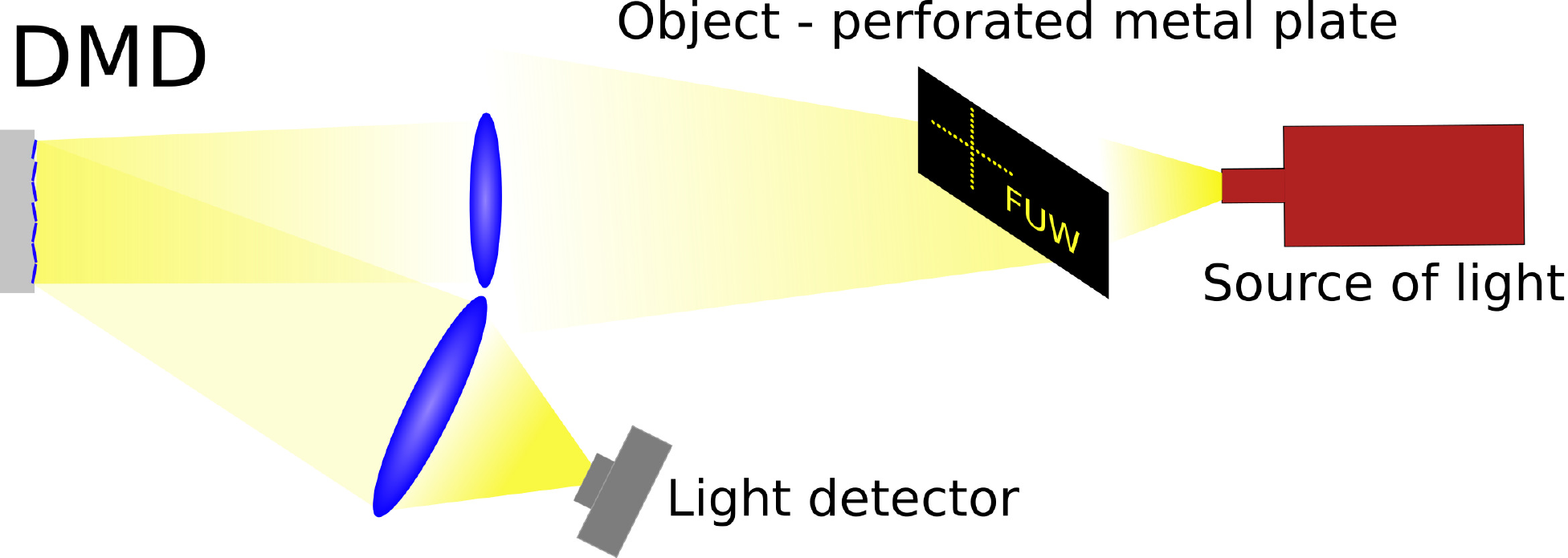}
\caption{Schematics of the optical SPI setup.}
\label{fig:SchemeI}
\end{figure}

We have validated the performance of the proposed MD-FDRI algorithm by implementing it in a classical SPI set-up with a DMD used at its full resolution. 
We were able to obtain high resolution SPI imaging of sparse objects. 
To this end we have prepared a translucent test pattern consisting of a set of holes in nontransparent metal foil with diameter in the range from 0.15mm to 0.41mm. The object is observed in transmission. The high contrast and brightness of the spots assure a high SNR of the measurement.  
Our experimental setup is depicted in Fig.~\ref{fig:SchemeI}. The object is illuminated by unpolarized light from a collimated LED source. Light transmitted through the object is collected by achromatic doublet and its image is formed on the DMD modulator (Vialux V-7001 XGA with DLP7000 chip). 
Compared to imaging at a lower resolution with the DMD pixels grouped into larger rectangular superpixels, here imaging of the object onto the DMD requires pixel-level precision. The DMD modulates the incident beam into two reflected beams. We will use only a single channel here but in fact the D-FDRI allows one to process the complementary channels using just a single reconstruction matrix $\mathbf{P}$~\cite{Pastuszczak_2021}. Then, the channels could be used for parallel detection through different spectral or polarization filters~\cite{Pastuszczak_2021}. 

The modulator is operated at $22.7$~kHz and at a spatial resolution of $1024\times 768$. We use a VIS-NIR amplified photodiode (Thorlabs PDA100A2) as the light detector. The signals are digitized with a digital oscilloscope (PicoScope 5000) at a sampling rate of $1/(256$~ns) and streamed through a USB bus for real-time processing.

\begin{figure}[!th]
\centering\includegraphics[width=14cm]{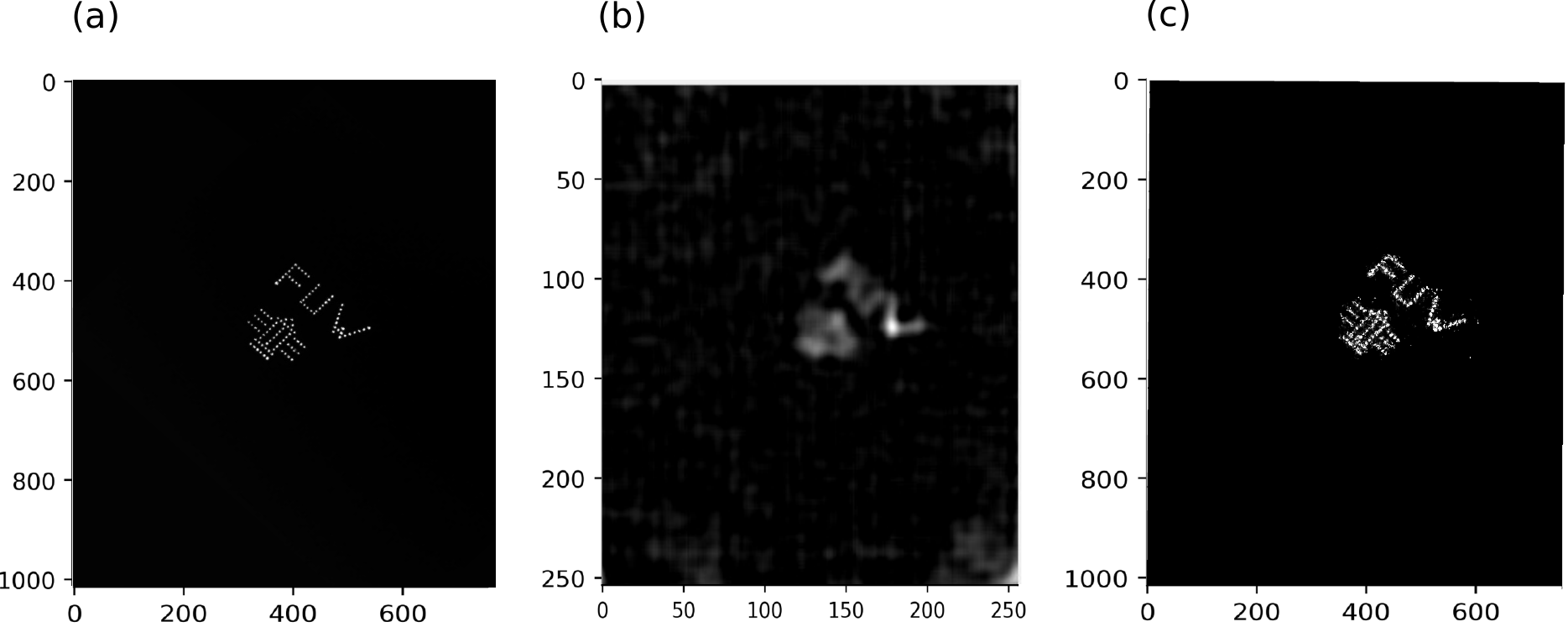}
\caption{Optical SPI measurement (a) Ground truth captured on 1278x1020 pixel camera. (b) 256x256 pixel image sampled with binarized DCT patterns reconstructed with method described in~\cite{Pastuszczak_2021} (compression ratio 0.03). (c) 1024x768 pixel image reconstructed with the proposed MD-FDRI algorithm (compression ratio 0.004).}
\label{fig:rekonstrukcja}
\end{figure}

\autoref{fig:rekonstrukcja} illustrates the performance of MD-FDRI  method applied to the high-resolution optical SPI imaging. \autoref{fig:rekonstrukcja}(a) depicts the ground truth image of the sparse test object captured with a 1278x1020 camera. The difference in the brightness of the subsequent spots comes from the different sizes of subsequent holes. 
\autoref{fig:rekonstrukcja}(c) shows the MD-FDRI reconstruction of the same image measured with our optical SPI setup using map-based sampling at full resolution of the DMD and with compression ratio of $0.4\%$. Despite the extreme compression, the reconstruction retains most of the details of the original object and high contrast. For comparison, in Fig.~\ref{fig:rekonstrukcja} (b) we present the same image measured using binarized DCT sampling patterns at the resolution $256 \times 256$ and compression ratio of $3 \%$, which is one of the commonly used SPI state-of-the-art sampling protocols. While in both cases the time needed for image acquisition is comparable, the reconstruction obtained with MD-FDRI provides significantly higher quality and accuracy in recovering image details.

\section{Conclusions}
In this paper we have demonstrated that using a classical DMD-based single-pixel imaging (SPI) optical set-up with non-adaptive binary sampling one may be capable of capturing and reconstructing sparse images at a high resolution within a fraction of a second. This is important as after over a decade of research on SPI, the reported set-ups hardly ever exceeded the resolution of $256\times 256$ falling far behind current resolution standards.

To this point we have proposed a novel differential binary high-resolution (n=$1024\times768$) highly compressive sampling scheme combined with an image reconstruction algorithm with a numerical cost of $O(n\cdot k)$. Here $n$ is the number of the DMD pixels and is equal to the resolution of the reconstructed images, and $k\approx 0.4\%\cdot n$ is the number of sampling functions, the acquisition of a single image takes $0.14$s, and a typical image reconstruction time is $t\approx 0.3$s on a Intel i9-10900X 3.70GHz CPU. The same algorithm is capable of reconstructing non-sparse images with reduced quality due to the strong compression. This work paves the way for practical high resolution SPI applications without more sophisticated modulation or detection elements than just a DMD and a photodiode. Additionally, we point a way to vary the SPI imaging quality using a simple diaphragm. Besides the usefulness of this property for instance for SPI microscopy, we would like to point that the trade off between resolution and field of view is something common to various imaging techniques involving an information channel with a limited bandwidth. A similar trade-off is known in classical microscopy as well. By limiting the field of view and decreasing the information content within the measured image it is possible to improve SPI imaging quality up to the full resolution of the DMD without any further adjustments to the measurement set-up or SPI algorithm.

\begin{backmatter}

\bmsection{Funding}
National Science Center, Poland - (RS,PW,RK)-UMO-2017/27/B/ST7/00885, (AP)-UMO-2019/35/D/ST7/03781.

\bmsection{Data Availability Statement}
Source data and source code will be provided by the authors at a reasonable request.

\bmsection{Disclosures}
The authors declare no conflicts of interest. 
\end{backmatter}

\section{References}

\bibliography{sample}

\begin{thebibliography}{10}
\newcommand{\enquote}[1]{``#1''}

\bibitem{Duarte_2008}
M.~F. Duarte, M.~A. Davenport, D.~Takhar, J.~N. Laska, T.~Sun, K.~F. Kelly, and
  R.~G. Baraniuk, \enquote{{Single-pixel imaging via compressive sampling},}
  {\protect\JournalTitle{{IEEE} Sig. Proc. Mag.}} \textbf{25}, 83--91 (2008).

\bibitem{Cand_s_2006}
E.~J. Cand{\`{e}}s, J.~K. Romberg, and T.~Tao, \enquote{{Stable signal recovery
  from incomplete and inaccurate measurements},}
  {\protect\JournalTitle{Communications on Pure and Applied Mathematics}}
  \textbf{59}, 1207--1223 (2006).

\bibitem{Gibson2020}
G.~M. Gibson, S.~D. Johnson, and M.~J. Padgett, \enquote{{Single-pixel imaging
  12 years on: a review},} {\protect\JournalTitle{Opt. Express}} \textbf{28},
  28190--28208 (2020).

\bibitem{Edgar_2018}
M.~P. Edgar, G.~M. Gibson, and M.~J. Padgett, \enquote{{Principles and
  prospects for single-pixel imaging},} {\protect\JournalTitle{Nat. Photonics}}
  \textbf{13}, 13--20 (2018).

\bibitem{Czajkowski_2018}
K.~M. Czajkowski, A.~Pastuszczak, and R.~Koty{\'{n}}ski, \enquote{{Real-time
  single-pixel video imaging with Fourier domain regularization},}
  {\protect\JournalTitle{Opt. Express}} \textbf{26}, 20009--20022 (2018).

\bibitem{Stantchev2020}
R.~Stantchev, X.~Yu, T.~Blu, and E.~Pickwell-MacPherson, \enquote{{Real-time
  terahertz imaging with a single-pixel detector},} {\protect\JournalTitle{Nat.
  Commun.}} \textbf{11}, 2535 (2020).

\bibitem{Higham2018}
C.~Higham, R.~Murray-Smith, M.~Padgett, and M.~Edgar, \enquote{{Deep learning
  for real-time single-pixel video},} {\protect\JournalTitle{Sci. Rep.}}
  \textbf{8}, 2369 (2018).

\bibitem{Rizvi2020}
S.~Rizvi, J.~Cao, K.~Zhang, and Q.~Hao, \enquote{{DeepGhost: real-time
  computational ghost imaging via deep learning},} {\protect\JournalTitle{Sci.
  Rep.}} \textbf{10}, 11400 (2020).

\bibitem{Wang:21}
Z.~Wang, W.~Zhao, A.~Zhai, P.~He, and D.~Wang, \enquote{{DQN} based
  single-pixel imaging,} {\protect\JournalTitle{Opt. Express}} \textbf{29},
  15463--15477 (2021).

\bibitem{Salvador-Balaguer2018}
E.~Salvador-Balaguer, P.~Latorre-Carmona, C.~Chabert, F.~Pla, J.~Lancis, and
  E.~Tajahuerce, \enquote{{Low-cost single-pixel 3D imaging by using an LED
  array},} {\protect\JournalTitle{Opt. Express}} \textbf{26}, 15623--15631
  (2018).

\bibitem{Xu2018}
Z.~Xu, W.~Chen, J.~Penuelas, M.~Padgett, and M.~Sun, \enquote{{1000 fps
  computational ghost imaging using LED-based structured illumination},}
  {\protect\JournalTitle{Op. Express}} \textbf{26}, 2427--2434 (2018).

\bibitem{Wang_2020}
M.~Wang, M.-J. Sun, and C.~Huang, \enquote{{Single-pixel 3D reconstruction via
  a high-speed {LED} array},} {\protect\JournalTitle{J.Phys. Photonics}}
  \textbf{2}, 025006 (2020).

\bibitem{Zhao_2019}
W.~Zhao, H.~Chen, Y.~Yuan, H.~Zheng, J.~Liu, Z.~Xu, and Y.~Zhou,
  \enquote{{Ultrahigh-Speed Color Imaging with Single-Pixel Detectors at Low
  Light Level},} {\protect\JournalTitle{Phys. Rev. Appl.}} \textbf{12}, 034049
  (2019).

\bibitem{Guerboukha_2018}
H.~Guerboukha, K.~Nallappan, and M.~Skorobogatiy, \enquote{{Toward real-time
  terahertz imaging},} {\protect\JournalTitle{Adv. Opt. Photon.}} \textbf{10},
  843--938 (2018).

\bibitem{Chen2019}
S.~Chen, L.~Du, K.~Meng, J.~Li, Z.~Zhai, Q.~Shi, Z.~Li, and L.~Zhu,
  \enquote{{Terahertz wave near-field compressive imaging with a spatial
  resolution of over $\lambda$/100},} {\protect\JournalTitle{Opt. Lett.}}
  \textbf{44}, 21--24 (2019).

\bibitem{Hahamovich:2021}
E.~Hahamovich, S.~Monin, Y.~Hazan, and A.~Rosenthal, \enquote{{Single pixel
  imaging at megahertz switching rates via cyclic Hadamard masks},}
  {\protect\JournalTitle{Nat. Commun.}} \textbf{12}, 4561 (2021).

\bibitem{Liu_oe_26_10048_2018}
C.~Liu, J.~Chen, J.~Liu, and X.~Han, \enquote{High frame-rate computational
  ghost imaging system using an optical fiber phased array and a low-pixel apd
  array,} {\protect\JournalTitle{Opt. Express}} \textbf{26}, 10048--10064
  (2018).

\bibitem{W.Yuwang2017}
{W. Yuwang}, {L. Yang}, {S. Jinli}, {S. Guohai}, {Q.Chang}, and {D. Qionghai},
  \enquote{{High Speed Computational Ghost Imaging via Spatial Sweeping},}
  {\protect\JournalTitle{Sci. Rep.}} \textbf{7}, 45325 (2017).

\bibitem{Soldevila:21}
F.~Soldevila, A.~J.~M. Lenz, A.~Ghezzi, A.~Farina, C.~D'Andrea, and
  E.~Tajahuerce, \enquote{Giga-voxel multidimensional fluorescence imaging
  combining single-pixel detection and data fusion,}
  {\protect\JournalTitle{Opt. Lett.}} \textbf{46}, 4312--4315 (2021).

\bibitem{Ghezzi:21}
A.~Ghezzi, A.~Farina, A.~Bassi, G.~Valentini, I.~Labanca, G.~Acconcia, I.~Rech,
  and C.~D'Andrea, \enquote{Multispectral compressive fluorescence lifetime
  imaging microscopy with a spad array detector,} {\protect\JournalTitle{Opt.
  Lett.}} \textbf{46}, 1353--1356 (2021).

\bibitem{Ke_2012}
J.~Ke and E.~Y. Lam, \enquote{{Object reconstruction in block-based compressive
  imaging},} {\protect\JournalTitle{Opt. Express}} \textbf{20}, 22102--22117
  (2012).

\bibitem{Mahalanobis_2014}
A.~Mahalanobis, R.~Shilling, R.~Murphy, and R.~Muise, \enquote{{Recent results
  of medium wave infrared compressive sensing},} {\protect\JournalTitle{Appl.
  Opt.}} \textbf{53}, 8060--8070 (2014).

\bibitem{Wu_2019}
Z.~Wu and X.~Wang, \enquote{{Focal plane array-based compressive imaging in
  medium wave infrared: modeling implementation, and challenges},}
  {\protect\JournalTitle{Appl. Opt.}} \textbf{58}, 8433--8441 (2019).

\bibitem{Stuart:IEEETCI-2021}
S.~Bennett, Y.~Noblet, P.~F. Griffin, P.~Murray, S.~Marshall, J.~Jeffers, and
  D.~Oi, \enquote{Compressive sampling using a pushframe camera,}
  {\protect\JournalTitle{IEEE Transactions on Computational Imaging}}
  \textbf{7}, 1069--1079 (2021).

\bibitem{Phillips:17}
D.~B. Phillips, M.-J. Sun, J.~M. Taylor, M.~P. Edgar, S.~M. Barnett, G.~M.
  Gibson, and M.~J. Padgett, \enquote{Adaptive foveated single-pixel imaging
  with dynamic supersampling,} {\protect\JournalTitle{Science Advances}}
  \textbf{3}, e1601782 (2017).

\bibitem{Qian:19}
Y.~Qian, R.~He, Q.~Chen, G.~Gu, F.~Shi, and W.~Zhang, \enquote{Adaptive
  compressed 3d ghost imaging based on the variation of surface normals,}
  {\protect\JournalTitle{Opt. Express}} \textbf{27}, 27862--27872 (2019).

\bibitem{Jiang:17}
H.~Jiang, S.~Zhu, H.~Zhao, B.~Xu, and X.~Li, \enquote{Adaptive regional
  single-pixel imaging based on the {Fourier} slice theorem,}
  {\protect\JournalTitle{Opt. Express}} \textbf{25}, 15118--15130 (2017).

\bibitem{He:21}
R.~He, Z.~Weng, Y.~Zhang, C.~Qin, J.~Zhang, Q.~Chen, and W.~Zhang,
  \enquote{Adaptive {Fourier} single pixel imaging based on the radial
  correlation in the {Fourier} domain,} {\protect\JournalTitle{Opt. Express}}
  \textbf{29}, 36021--36037 (2021).

\bibitem{Pastuszczak_2021}
A.~Pastuszczak, R.~Stojek, P.~Wr{\'o}bel, and R.~Koty{\'{n}}ski,
  \enquote{{Differential real-time single-pixel imaging with Fourier domain
  regularization - applications to VIS-IR imaging and polarization imaging},}
  {\protect\JournalTitle{Opt. Express}} \textbf{29}, 2685--26700 (2021).

\bibitem{Radwell_2014}
N.~Radwell, K.~J. Mitchell, G.~M. Gibson, M.~P. Edgar, R.~Bowman, and M.~J.
  Padgett, \enquote{{Single-pixel infrared and visible microscope},}
  {\protect\JournalTitle{Optica}} \textbf{1}, 285--289 (2014).

\bibitem{Zhang2015}
Z.~Zhang, X.~Ma, and J.~Zhong, \enquote{{Single-pixel imaging by means of
  Fourier spectrum acquisition},} {\protect\JournalTitle{Nat. Commun.}}
  \textbf{6}, 6225 (2015).

\bibitem{Czajkowski_2019}
K.~M. Czajkowski, A.~Pastuszczak, and R.~Koty{\'{n}}ski, \enquote{{Single-pixel
  imaging with sampling distributed over simplex vertices},}
  {\protect\JournalTitle{Opt. Lett.}} \textbf{44}, 1241--1244 (2019).

\bibitem{DFDRI:code}
A.~Pastuszczak, R.~Stojek, P.~Wr{\'o}bel, and R.~Koty{\'{n}}ski,
  \enquote{Differential {Fourier} domain regularized inversion (d-fdri source
  code),}  (2021). Github, \url{http://www.github.com/rkotynski/D_FDRI}.

\bibitem{Sun_2018}
M.-J. Sun, Z.-H. Xu, and L.-A. Wu, \enquote{{Collective noise model for focal
  plane modulated single-pixel imaging},} {\protect\JournalTitle{Opt. Lasers
  Eng.}} \textbf{100}, 18--22 (2018).

\bibitem{Lukosz:JOSA-1967-57-932}
W.~Lukosz, \enquote{Optical systems with resolving powers exceeding the
  classical limit. {II},} {\protect\JournalTitle{J. Opt. Soc. Am.}}
  \textbf{57}, 932--941 (1967).

\bibitem{Sheppard:Micron-2007-38-165}
C.~J. Sheppard, \enquote{Fundamentals of superresolution,}
  {\protect\JournalTitle{Micron}} \textbf{38}, 165--169 (2007).

\end{thebibliography}

\clearpage

\renewcommand{\thefigure}{S\arabic{figure}}
\setcounter{figure}{0}
\renewcommand{\thesection}{S\arabic{section}}
\setcounter{section}{0}

\definecolor{codegreen}{rgb}{0,0.6,0}
\definecolor{codegray}{rgb}{0.5,0.5,0.5}
\definecolor{codepurple}{rgb}{0.58,0,0.82}
\definecolor{backcolour}{rgb}{0.95,0.95,0.92}

\lstdefinestyle{mystyle}{
	backgroundcolor=\color{backcolour},   
	commentstyle=\color{codegreen},
	keywordstyle=\color{magenta},
	numberstyle=\tiny\color{codegray},
	stringstyle=\color{codepurple},
	basicstyle=\ttfamily\footnotesize,
	breakatwhitespace=false,         
	breaklines=true,                 
	captionpos=b,                    
	keepspaces=true,                 
	numbers=left,                    
	numbersep=5pt,                  
	showspaces=false,                
	showstringspaces=false,
	showtabs=false,                  
	tabsize=2
}
\lstset{style=mystyle}

\title{Supplementary Materials: Single pixel imaging at high pixel resolutions}

	\author{Rafa{\l} Stojek,\authormark{1,2} Anna Pastuszczak,\authormark{1} Piotr Wr{\'o}bel,\authormark{1} and Rafa{\l} Koty{\'n}ski \authormark{1,*}}
	
	\address{\authormark{1}University of Warsaw, Faculty of Physics, Pasteura 5, 02-093 Warsaw, Poland\\
		\authormark{2}Vigo System, Poznańska 129/133, 05-850 Ożarów Mazowiecki, Poland\\}
	
	\email{\authormark{*}Rafal.Kotynski@fuw.edu.pl} 



\section{Introduction}
This document contains supplementary materials to our Optics Express paper entitled "Single pixel imaging at high pixel resolutions" which we will further call the main paper. Supplementary materials provide more detailed information on the generation of image maps (see Section \ref{sect:maps}), on constructing the measurement matrix from the image maps and removing linear dependence from the sampling patterns (see Section \ref{sect:encod}), and on the resolution difference between the  Fourier domain regularized inversion (FDRI) and image map-based Fourier domain regularized inversion (MD-FDRI which is introduced in the main paper) methods (see Section \ref{sect:resolution}). In the last section \ref{sect:atlas} an atlas of the lookup tables used for encoding maps with different number of image regions is given and sample SPI results obtained with varying parameter values are compared.

	\section{Calculation of image maps and of the measurement matrix\label{sect:maps}}
\begin{figure}[!ht]
	\includegraphics[width=12cm]{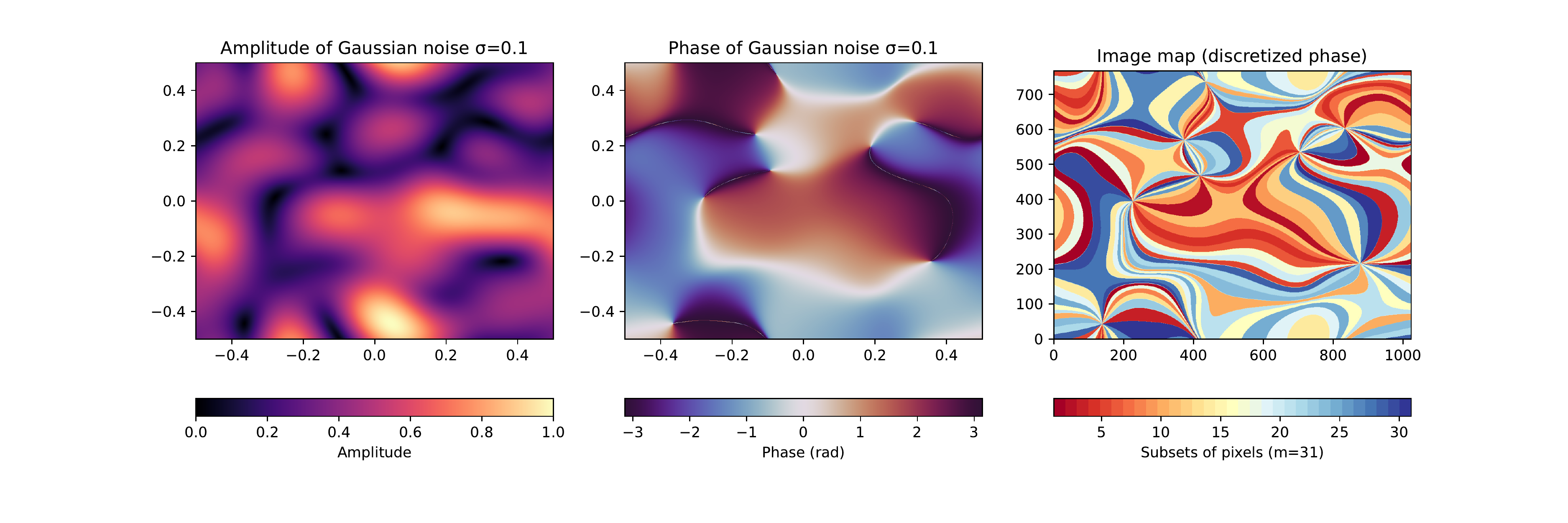}
	\includegraphics[width=12cm]{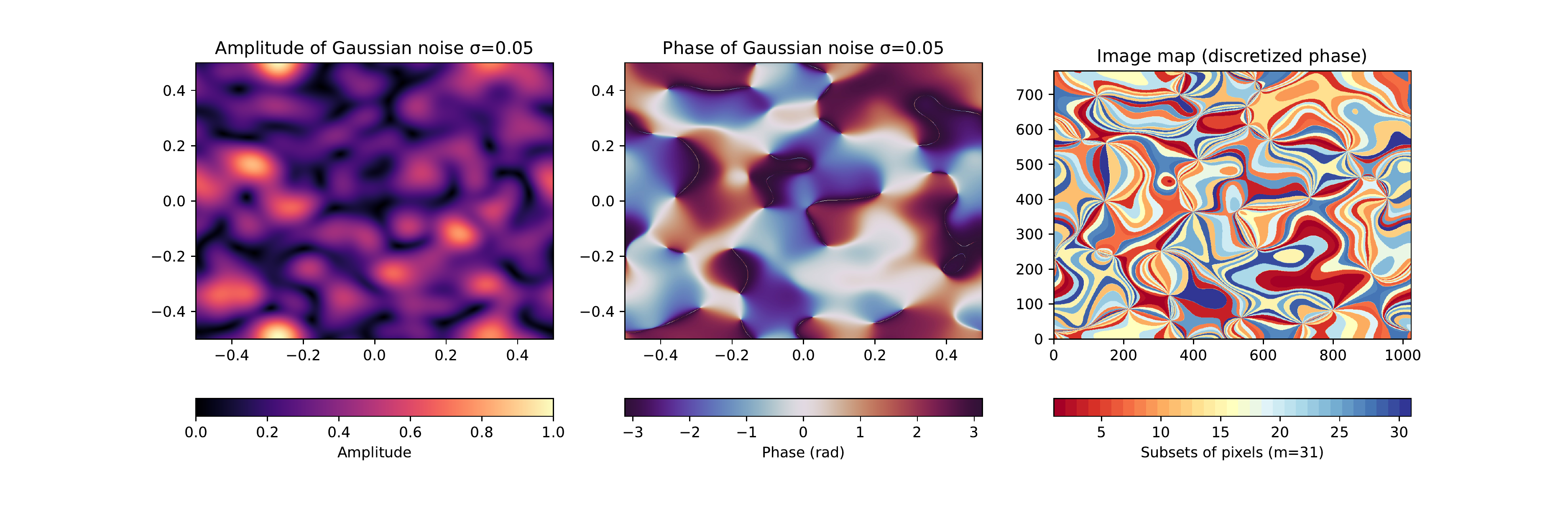}
	\includegraphics[width=12cm]{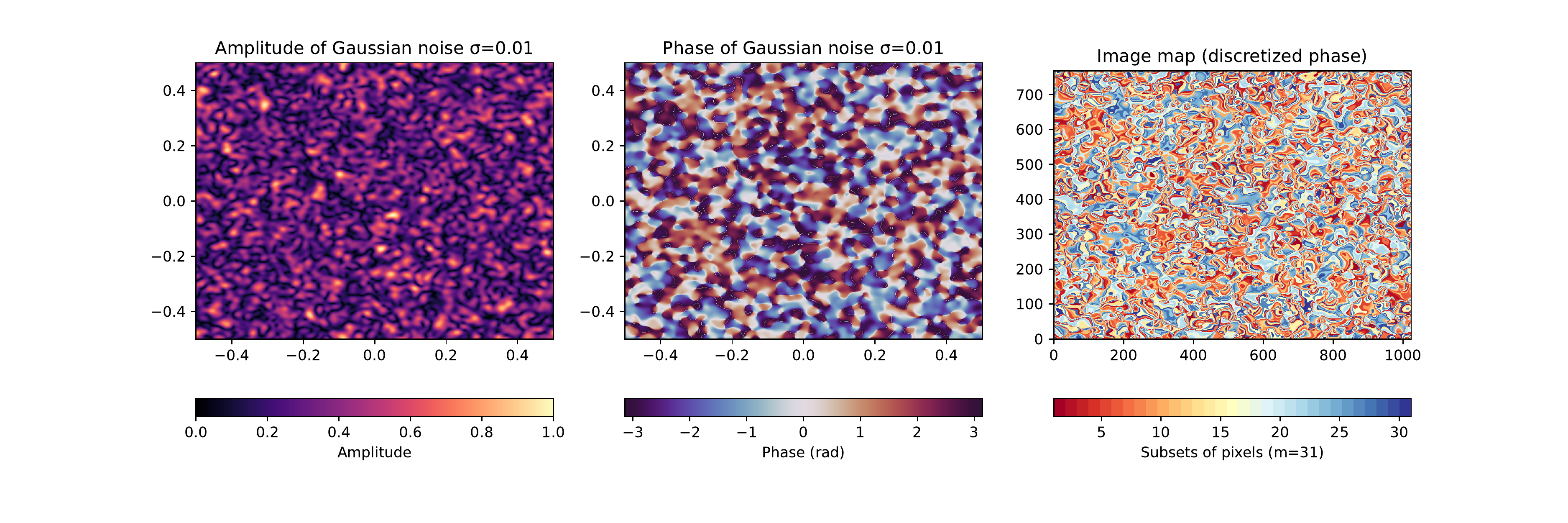}
	\caption{Sample image maps: the left and middle columns show the amplitude and the phase of Gaussian noise samples; the right column shows the phase discretized into $m=31$ levels. For better visualisation, the color assigned to various phase ranges is randomized.}\label{fig.s1.maps}
\end{figure}

	In the main paper, the binary patterns included in the rows of the measurement matrix $\mathbf{M}$ play a double role. First, the sampling provides the usual
information about the spatial contents of the measured image. On top of this it should give some
guarantees about the possibility of identifying empty regions of the image. With this aim, sampling patterns are generated with the help of auxiliary image maps $\mathbf{m}_1, \mathbf{m}_2,... \mathbf{m}_{l}$ whose role is discussed in Sect. 2.3 of the main paper. Every map defines a partitioning of all image pixels into $m$ groups . The first map is used to create $m+1$ sampling patterns, and every other map is used to create $m-1$ patterns. The particular form of the maps used by us is rather arbitrary. It could be further optimized but image maps very different from those proposed in the main paper could also work fine with the MD-FDRI algorithm. This is why in the main paper we have not paid too much focus on the way we generate the maps. In short, the maps are obtained by generating spatially correlated Gaussian complex-valued noise and by assigning the label $j\in\lbrace 1,m \rbrace$ based on the uniform quantization of the phase level of a noise realization.  The precise procedure is defined by the Python program shown in Listing~1. Sample image maps generated by this program are shown in Fig.~\ref{fig.s1.maps}

\begin{lstlisting}[language=Python, caption=Python function for generating image maps] 
import numpy as np
def create_image_map(dim=(768, 1024), m=31):
    '''
	Create an image map with integer values in range [1,m] based on the uniform
	quantization of the phase level of spatially correlated 
	Gaussian complex-valued noise 
	
	Parameters
	----------
	dim : dimensions of the image map. The default is (768,1024).
	m : number of pixel regions in the map. The default is 31.
	sgm: correlation width. The default is max(np.random.random()*.1, 0.004)
	Returns: an image map with integer values in range [1,m]
	'''
	sgm = 0.01 + 0.99 * np.random.rand()**2 # random value in range [.01,.1]	
	x, y = np.meshgrid(np.linspace(-.5,.5, dim[1]),np.linspace(-.5, .5, dim[0]))
	# Generate complex correlated Gaussian zero-mean noise
	u = np.fft.fft2(np.fft.fftshift(np.exp(-x**2/(2*sgm**2)-y**2/(2*sgm**2))))
	u[0, 0] = 0
	u = np.fft.ifft2(np.fft.fft2(np.random.randn(*dim) +1j * np.random.randn(*dim))*u)
	# Find the phase of the Gaussian noise
	new_map = np.angle(u).reshape(-1)
	# Discretize the phase into m levels
	new_map[np.argsort(new_map)] = (m*np.arange(new_map.size))//new_map.size
	new_map = np.array(new_map.reshape(dim), dtype=np.uint8)
	# randomly permute the pixel regions  (for better visualisation only)
	new_map = 1+np.random.permutation(m)[new_map]
	return new_map  # return a map of shape dim with values [1..m]
\end{lstlisting}

The measurement matrix $\mathbf{M}$ is composed of rows that represent binary sampling patterns. These sampling patterns are obtained by processing the list of image maps.
Translation of image maps into binary patterns is performed with the help of subsequent rows of an auxiliary binary matrix $\mathbf{A}$ which play the role
of lookup tables. This procedure is described in Sect. 2.3 and in Fig. 1. of the main paper. The following Python function may be used to build a list of image maps and to calculate the measurement matrix:
\begin{lstlisting}[language=Python, caption=Python function for generating the measurement matrix from the image maps and a lookup table]
def create_measurement_matrix(A, dim=(768, 1024),l=100):
	'''
	Create the binary measurement matrix M with rows representing sampling patterns
	
	Parameters
	----------
	A : lookup table
	dim : dimension of image maps. The default is (768, 1024).
	l : number of image maps. The default is 100.
	
	Returns
	-------
	M : the binary measurement matrix
	image_maps :  an array with image maps
	'''
	m=A.shape[1]
	M=np.zeros((m+1+(l-1)*(m-1),np.prod(dim)),dtype=np.uint8)
	image_maps=np.array([create_image_map(m=m,dim=dim) for _ in range(l)])
	row=0
	for i in range(l):# loop over image maps
		for j in range(m):# loop over regions in the map
			M[row:row+A.shape[0], image_maps[i].reshape(-1)==j+1]=A[:,j].reshape((-1,1))
		row+=A.shape[0]
		if i==0:
			A=A[2:,:] # first two rows are used with the first image map only
	return M, image_maps # return the measurement matrix and image maps	
\end{lstlisting}

\section{The lookup table - a differential binary code for encoding binary patterns}\label{sect:encod}
Here we overview the properties of the lookup table $\mathbf{A}$ that we use to translate the image maps into sampling patterns. $\mathbf{A}$ is a binary $m+1\times m$ matrix. The main property that $\mathbf{A}$ should fulfill is that $\mathbf{D}\cdot\mathbf{A}$ is a full-rank matrix, i.e.  $rank(\mathbf{D}\cdot\mathbf{A})=m$. Given an image map, subsequent rows of $\mathbf{A}$ are used to construct a sequence of binary sampling patterns. The columns of $\mathbf{A}$  are linked to regions of pixels of that image map. Since $\mathbf{D}\cdot\mathbf{A}$ is a full-rank matrix, the differential measurement with the so created sampling patterns is equivalent to a measurement with patterns defined by separate pixel regions of the map. Yet, the proposed approach is better in terms of light efficiency and it is differential so a constant detector bias signal (e.g. from the dark current) is automatically disregarded in the measurement. By keeping similar surfaces of pixel regions within a map and by choosing matrices $\mathbf{A}$ with a similar number of ones and zeros in every row we assure that the SPI detection signal does not experience large variations. This comes in contrast to SPI schemes in which the mean value of image is measured with all DMD pixels in one position while the other patterns consist of approximately half of pixels in one position, a situation which leads to unnecessarily high requirements for the bit-depth of the DAQ.

We note that any two matrices $\mathbf{A}$ with permuted columns may be considered equivalent because the columns are arbitrarily assigned to pixel regions of an image map. We also note that when two or more image maps are used in a sequence to produce the corresponding sampling patterns, one could easily end up with measuring redundant information. One origin for this redundancy is that all the regions of any two maps cover exactly the same area (namely all image pixels) and for instance the image intensity could be measured with sampling patterns built from any map. The second is less obvious and results from the differential operator acting between the last and first sampling patterns from two subsequent maps. In effect to avoid redundancy only the first image map should be used to build $m+1$ sampling patterns and for each following map only $m-1$ patterns need to be created (per map). For this purpose we neglect the first two rows of $\mathbf{A}$ for all image maps but the first one. For this to work it is not sufficient to know that $rank(\mathbf{D}\cdot\mathbf{A})=m$ but also it is necessary to make sure that a block matrix composed of the last row of $\mathbf{A}$ as one block, and $\mathbf{A}$ with first two rows taken away as the second block, after applying the difference operator and adding a row consisting of $m$ ones and $m$ minus ones (representing an equation for the image intensities measured with two image maps) is a full rank matrix. This condition may be easier to formulate with a piece of Python code than with an equation: 
\begin{lstlisting}[language=Python, caption=Python function for generating the measurement matrix from the image maps and a lookup table]
def tst_lookup_table(A):
'''
Check if the lookup table A fullfills the required conditions

Parameters
----------
A : matrix A

Returns
-------
TYPE
True:ok, False: not ok
'''
	a=A.astype(float)
	m=a.shape[1]
	if a.shape[0]!=m+1:
		return False # matrix size not ok
	if np.linalg.matrix_rank(np.diff(a,axis=0))!=m:
		return False # D.A is not full rank
	b=np.vstack( (np.diff( np.hstack( (np.zeros((m-1,m)),a[2:,:])),axis=0), np.hstack( (np.ones((1,m)),-np.ones((1,m)))  ) ))
	tst=np.linalg.matrix_rank(b)==m-1 # is it ok to disregard first two rows of A for subsequent image maps?
	return tst	
\end{lstlisting}

\section{Resolution test}\label{sect:resolution}
Nonadaptive compressive imaging at a very low compression ratio (like the one considered here $k/n\approx0.4\%$) is challenging when the images are neither sparse in the spatial nor in the Fourier domains. The most common approach in optical SPI is to use sampling with a low spatial frequency subset of some basis functions (such as DCT, Fourier, Walsh-Hadamard basis etc.). In fact, even for sparse high resolution images their spatial spectra have usually the highest amplitudes at low spatial frequencies. On the other hand, retaining a high resolution is impossible without high spatial frequency information. 

\begin{figure}
	\includegraphics[width=14cm]{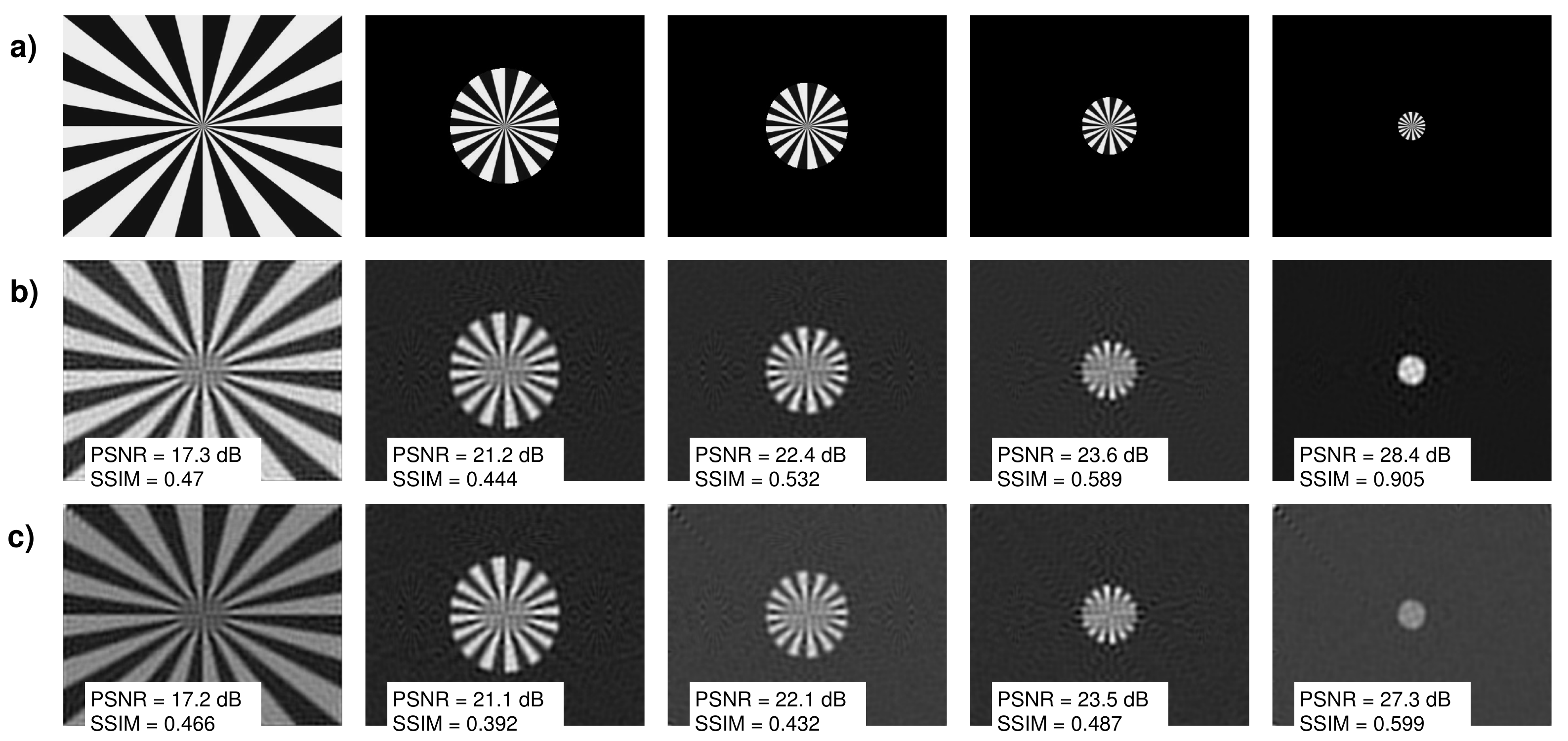}
	\caption{DCT-FDRI image reconstruction of the Siemens star resolution test - for comparisons with Fig.~5(a,c,e) of the main paper which shows similar results obtained with MD-FDRI obtained with the same number of sampling patterns ($k=3002$) and at the same resolution ($n=1024\times 768$). a)~ground truth; b)~noise-free measurement; c)~noisy measurement\label{fig:comp_fig5}}
\end{figure}

In Fig.~5 of the main paper we have shown MD-FDRI imaging results of a Siemens resolution test (with and without noise, and at the full or limited field of view). Here we compare these results with DCT-FDRI imaging (FDRI with binarized DCT sampling functions not based on image maps, and with a single step FDRI image reconstruction). The DCT-FDRI results are presented in Fig.~\ref{fig:comp_fig5}(b,c) which may be compared to Fig.~5(c,e) of the main paper. The DCT-FDRI performs better for an image with a full field of view both in a noise-free and noisy situations. This is confirmed by the values of PSNR and SSIM criteria as well as by visual examination. However, the center of the resolution test appears blurred, also when the field of view is limited. Therefore DCT-FDRI cannot be called a high-resolution method. On the other hand, for all the situations with a limited field of view, the MD-FDRI performs better in terms of PSNR and SSIM. Also a visual examination allows to appreciate the resolution obtained within the central part of the Siemens star (See the right columns of Fig.~5(c,e) of the main paper).

\section{Atlas of binary lookup tables}\label{sect:atlas}
The optimal choice of the parameters $l$ and $m$ is highly object dependent. For gray-scale dense images, other SPI methods may give better results than MD-FDRI. For sparse images or images with a limited FOV, optimal $l$ and $m$ and kind of image maps depend on the image complexity and kind of features included in an image. In Figs.~\ref{fig:comp_l_m_a},~\ref{fig:comp_l_m_b}  we show sample image reconstructions with and without noise for different values of $l$ an $m$ taken in such a way that the total number of sampling patterns $k$ is approximately constant. The results from this comparison indicate that the reconstruction quality does not vary in a simple or regular way with $l$ and $m$.
In the end of this section we provide a list of binary lookup tables of different sizes (for odd values of $m$ up to $m=31$). We note that these matrices are not unique.  

From our experience, usually the larger is $m$, the better for the SPI imaging quality, but the results vary depending on the composition of sparse images.
The lookup table used by us in this work is the largest we have calculated (with brute force binary search that becomes practically impossible to continue for larger $m$).    

\begin{landscape}
\begin{figure}
	\includegraphics[height=10cm]{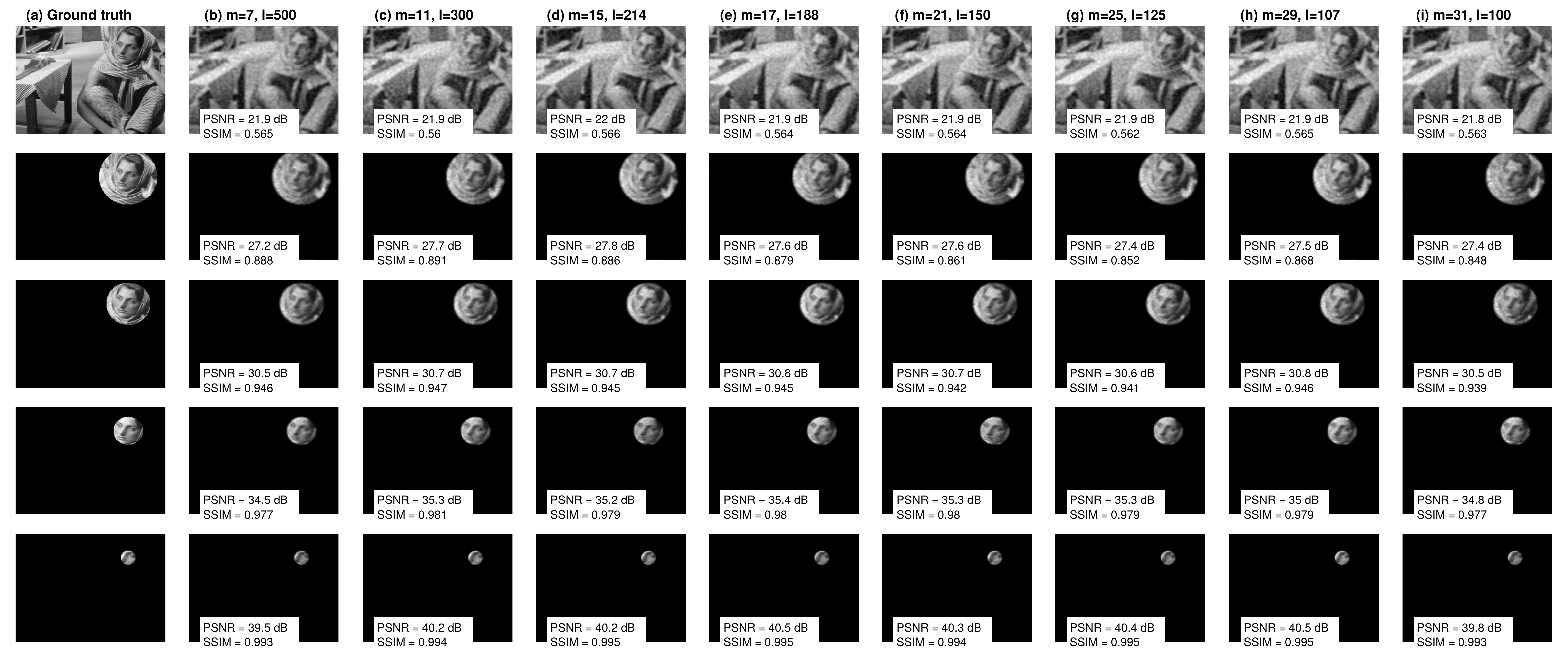}
	\caption{Comparison of MD-FDRI image reconstructions obtained for different values of $l$ and $m$ (with approximately preserved number of sampling patterns $k$) (noise free measurement). The comparison indicates that there is no simple or regular dependence between the image reconstruction quality and the parameters $l$ and $m$. The result is in fact largely object dependent.\label{fig:comp_l_m_a}}
\end{figure}
\begin{figure}
	\includegraphics[height=10cm]{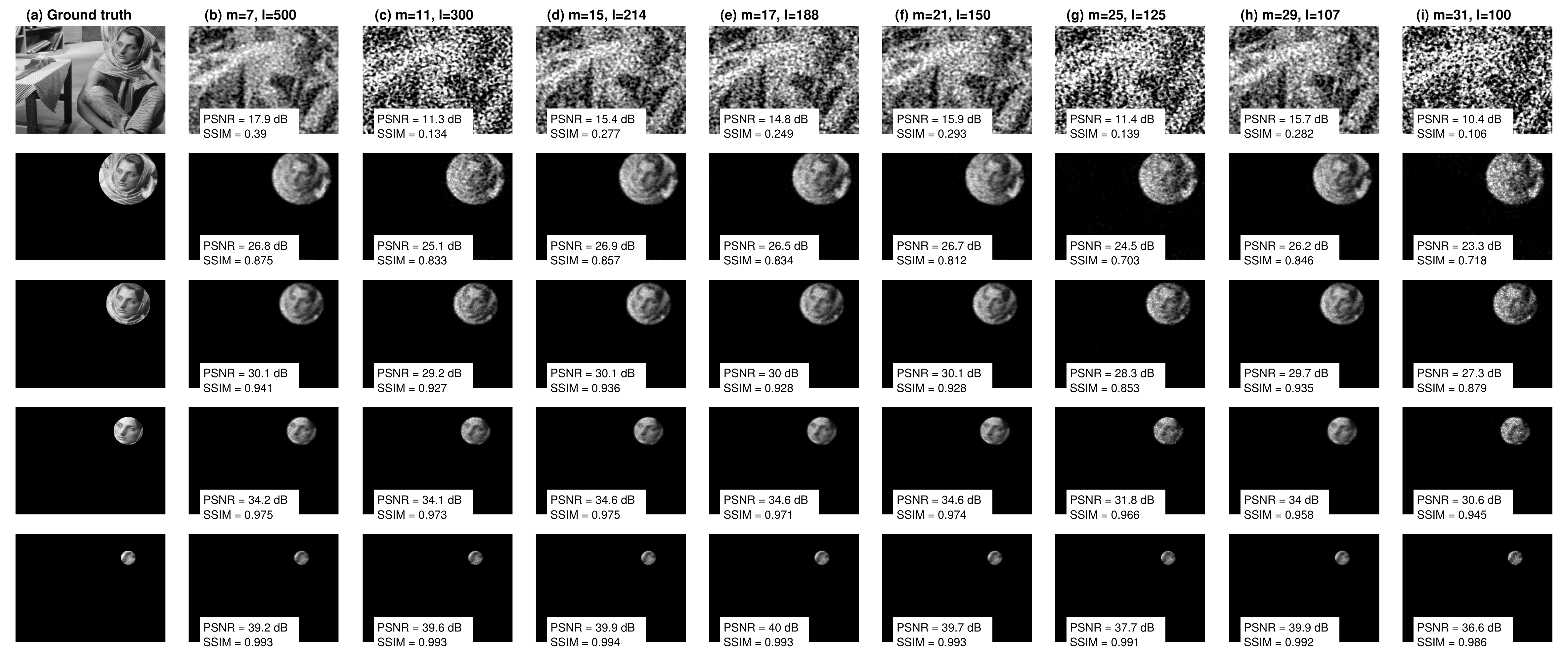}
	\caption{Comparison of MD-FDRI image reconstructions obtained for different values of $l$ and $m$ (with approximately preserved number of sampling patterns $k$) (noisy measurement, $\sigma=5\cdot 10^{-4}$). The comparison indicates that there is no simple or regular dependence between the image reconstruction quality and the parameters $l$ and $m$. The result is in fact largely object dependent.\label{fig:comp_l_m_b}}
\end{figure}
\end{landscape}

\begin{eqnarray*}
	\mathbf{A}_{m=3}=\scalemath{0.5}{\left[\begin{matrix}0 & 1 & 1\\1 & 0 & 0\\0 & 1 & 0\\1 & 1 & 0\end{matrix}\right]}\\
	\mathbf{A}_{m=5}=\scalemath{0.5}{\left[\begin{matrix}0 & 0 & 1 & 1 & 1\\1 & 0 & 1 & 0 & 1\\0 & 0 & 1 & 0 & 1\\1 & 1 & 1 & 0 & 0\\0 & 1 & 0 & 1 & 0\\1 & 0 & 1 & 0 & 0\end{matrix}\right]}\\
	\mathbf{A}_{m=7}=\scalemath{0.5}{\left[\begin{matrix}0 & 0 & 0 & 1 & 1 & 1 & 1\\1 & 1 & 1 & 0 & 0 & 1 & 0\\0 & 0 & 1 & 0 & 1 & 0 & 1\\1 & 0 & 0 & 0 & 1 & 1 & 0\\0 & 0 & 1 & 1 & 0 & 0 & 1\\1 & 0 & 1 & 0 & 1 & 0 & 1\\1 & 1 & 0 & 0 & 0 & 1 & 0\\0 & 0 & 1 & 1 & 1 & 0 & 1\end{matrix}\right]}\\
	\mathbf{A}_{m=9}=\scalemath{0.5}{\left[\begin{matrix}0 & 0 & 0 & 0 & 1 & 1 & 1 & 1 & 1\\0 & 1 & 1 & 1 & 1 & 0 & 1 & 0 & 0\\0 & 0 & 1 & 0 & 1 & 1 & 0 & 1 & 0\\0 & 1 & 1 & 0 & 1 & 1 & 0 & 1 & 0\\1 & 0 & 0 & 0 & 1 & 1 & 0 & 1 & 1\\1 & 1 & 1 & 1 & 0 & 1 & 0 & 0 & 0\\1 & 1 & 0 & 0 & 1 & 1 & 0 & 0 & 0\\0 & 0 & 0 & 1 & 1 & 0 & 0 & 1 & 1\\0 & 1 & 0 & 0 & 0 & 0 & 1 & 1 & 1\\0 & 1 & 1 & 0 & 0 & 1 & 1 & 0 & 0\end{matrix}\right]}\\
	\mathbf{A}_{m=11}=\scalemath{0.5}{\left[\begin{array}{ccccccccccc}0 & 0 & 0 & 0 & 0 & 1 & 1 & 1 & 1 & 1 & 1\\0 & 0 & 0 & 1 & 1 & 0 & 1 & 0 & 0 & 1 & 1\\0 & 0 & 0 & 1 & 0 & 1 & 0 & 1 & 1 & 1 & 0\\0 & 0 & 1 & 1 & 1 & 1 & 0 & 0 & 1 & 0 & 1\\1 & 1 & 0 & 0 & 0 & 0 & 0 & 1 & 1 & 1 & 0\\1 & 0 & 1 & 1 & 0 & 1 & 0 & 0 & 0 & 0 & 1\\1 & 1 & 1 & 0 & 0 & 0 & 1 & 0 & 0 & 1 & 0\\0 & 0 & 0 & 1 & 1 & 1 & 0 & 1 & 0 & 0 & 1\\0 & 0 & 0 & 1 & 1 & 0 & 1 & 1 & 1 & 1 & 0\\1 & 1 & 1 & 1 & 0 & 0 & 0 & 1 & 0 & 0 & 1\\1 & 1 & 0 & 1 & 0 & 1 & 1 & 0 & 0 & 1 & 0\\1 & 0 & 1 & 1 & 1 & 0 & 0 & 0 & 0 & 1 & 1\end{array}\right]}\\
	\mathbf{A}_{m=13}=\scalemath{0.5}{\left[\begin{array}{ccccccccccccc}0 & 0 & 0 & 0 & 0 & 0 & 1 & 1 & 1 & 1 & 1 & 1 & 1\\1 & 0 & 0 & 1 & 0 & 1 & 0 & 1 & 1 & 0 & 0 & 1 & 1\\0 & 0 & 1 & 1 & 0 & 0 & 0 & 1 & 1 & 1 & 0 & 0 & 1\\1 & 0 & 0 & 0 & 0 & 0 & 1 & 1 & 1 & 0 & 0 & 1 & 1\\1 & 1 & 1 & 0 & 1 & 0 & 0 & 1 & 0 & 0 & 0 & 1 & 0\\0 & 1 & 0 & 0 & 1 & 1 & 0 & 0 & 1 & 0 & 1 & 1 & 1\\1 & 1 & 1 & 1 & 0 & 0 & 0 & 1 & 0 & 1 & 0 & 0 & 1\\1 & 0 & 0 & 0 & 0 & 1 & 0 & 0 & 0 & 1 & 1 & 1 & 1\\1 & 1 & 1 & 0 & 1 & 0 & 0 & 0 & 1 & 1 & 0 & 0 & 1\\0 & 1 & 1 & 1 & 1 & 0 & 0 & 1 & 1 & 0 & 0 & 0 & 0\\1 & 1 & 0 & 0 & 1 & 0 & 0 & 0 & 1 & 1 & 0 & 0 & 1\\0 & 0 & 0 & 1 & 0 & 1 & 1 & 1 & 1 & 0 & 0 & 0 & 1\\0 & 0 & 0 & 0 & 0 & 0 & 1 & 1 & 1 & 1 & 1 & 0 & 1\\1 & 0 & 0 & 0 & 1 & 1 & 1 & 1 & 0 & 0 & 1 & 1 & 0\end{array}\right]}\\
	\mathbf{A}_{m=15}=\scalemath{0.5}{\left[\begin{array}{ccccccccccccccc}0 & 0 & 0 & 0 & 0 & 0 & 0 & 1 & 1 & 1 & 1 & 1 & 1 & 1 & 1\\0 & 1 & 0 & 1 & 0 & 0 & 0 & 0 & 1 & 1 & 0 & 1 & 1 & 1 & 0\\1 & 1 & 0 & 1 & 1 & 0 & 0 & 0 & 1 & 0 & 1 & 1 & 0 & 1 & 0\\0 & 1 & 0 & 0 & 1 & 0 & 1 & 1 & 1 & 0 & 1 & 0 & 1 & 1 & 0\\0 & 0 & 1 & 1 & 1 & 0 & 0 & 1 & 0 & 1 & 1 & 0 & 1 & 0 & 0\\0 & 1 & 0 & 1 & 1 & 1 & 1 & 0 & 0 & 0 & 1 & 0 & 0 & 1 & 1\\1 & 0 & 1 & 0 & 1 & 0 & 0 & 1 & 0 & 0 & 0 & 1 & 0 & 1 & 1\\0 & 1 & 1 & 1 & 0 & 0 & 0 & 1 & 1 & 1 & 1 & 0 & 0 & 0 & 0\\0 & 1 & 1 & 1 & 0 & 0 & 0 & 1 & 0 & 0 & 0 & 1 & 1 & 1 & 1\\1 & 0 & 1 & 1 & 0 & 1 & 1 & 0 & 0 & 1 & 1 & 0 & 0 & 1 & 0\\0 & 1 & 0 & 1 & 0 & 0 & 1 & 1 & 0 & 0 & 1 & 0 & 0 & 1 & 1\\0 & 1 & 0 & 0 & 1 & 1 & 1 & 0 & 0 & 0 & 0 & 1 & 1 & 1 & 0\\1 & 0 & 1 & 0 & 0 & 0 & 1 & 0 & 1 & 0 & 1 & 0 & 1 & 0 & 1\\0 & 0 & 1 & 1 & 1 & 0 & 0 & 0 & 1 & 1 & 1 & 1 & 0 & 0 & 1\\1 & 1 & 1 & 0 & 0 & 0 & 1 & 1 & 1 & 1 & 0 & 0 & 1 & 0 & 0\\1 & 0 & 0 & 0 & 0 & 1 & 0 & 0 & 1 & 1 & 0 & 1 & 0 & 1 & 1\end{array}\right]}\\
	\mathbf{A}_{m=17}=\scalemath{0.5}{\left[\begin{array}{ccccccccccccccccc}0 & 0 & 0 & 0 & 0 & 0 & 0 & 0 & 1 & 1 & 1 & 1 & 1 & 1 & 1 & 1 & 1\\1 & 0 & 0 & 1 & 0 & 0 & 1 & 0 & 0 & 1 & 0 & 1 & 1 & 0 & 1 & 1 & 0\\1 & 1 & 1 & 0 & 1 & 0 & 0 & 0 & 0 & 0 & 1 & 0 & 1 & 1 & 0 & 0 & 1\\0 & 1 & 0 & 1 & 0 & 0 & 0 & 1 & 1 & 0 & 0 & 0 & 1 & 1 & 1 & 1 & 1\\1 & 0 & 0 & 1 & 0 & 0 & 0 & 1 & 1 & 0 & 1 & 1 & 1 & 1 & 0 & 1 & 0\\0 & 1 & 1 & 0 & 0 & 1 & 0 & 0 & 0 & 1 & 1 & 1 & 0 & 1 & 0 & 1 & 0\\0 & 1 & 0 & 1 & 0 & 0 & 0 & 1 & 1 & 1 & 0 & 1 & 0 & 0 & 1 & 0 & 1\\1 & 1 & 1 & 0 & 0 & 1 & 0 & 0 & 0 & 0 & 1 & 0 & 0 & 1 & 1 & 1 & 0\\0 & 0 & 0 & 1 & 0 & 1 & 0 & 1 & 1 & 1 & 0 & 0 & 1 & 1 & 0 & 0 & 1\\1 & 0 & 1 & 1 & 0 & 1 & 1 & 0 & 1 & 1 & 0 & 0 & 1 & 0 & 1 & 0 & 0\\1 & 1 & 1 & 1 & 1 & 1 & 0 & 0 & 0 & 0 & 1 & 0 & 0 & 1 & 0 & 0 & 1\\0 & 0 & 0 & 1 & 1 & 1 & 0 & 1 & 0 & 1 & 1 & 1 & 1 & 0 & 1 & 0 & 0\\1 & 1 & 0 & 1 & 0 & 0 & 1 & 0 & 0 & 1 & 0 & 1 & 1 & 0 & 1 & 1 & 0\\1 & 1 & 1 & 1 & 1 & 1 & 0 & 0 & 1 & 0 & 0 & 0 & 0 & 0 & 1 & 0 & 0\\1 & 1 & 1 & 0 & 0 & 1 & 0 & 1 & 0 & 1 & 0 & 1 & 1 & 0 & 0 & 1 & 0\\1 & 1 & 1 & 0 & 0 & 0 & 0 & 0 & 0 & 0 & 1 & 1 & 1 & 1 & 0 & 1 & 0\\0 & 0 & 0 & 0 & 1 & 0 & 1 & 0 & 0 & 0 & 1 & 1 & 1 & 0 & 1 & 1 & 1\\0 & 0 & 1 & 0 & 1 & 0 & 1 & 1 & 1 & 1 & 0 & 1 & 0 & 1 & 1 & 0 & 0\end{array}\right]}\\
	\mathbf{A}_{m=19}=\scalemath{0.5}{\left[\begin{array}{ccccccccccccccccccc}0 & 0 & 0 & 0 & 0 & 0 & 0 & 0 & 0 & 1 & 1 & 1 & 1 & 1 & 1 & 1 & 1 & 1 & 1\\1 & 0 & 0 & 1 & 0 & 0 & 1 & 1 & 0 & 1 & 1 & 0 & 0 & 1 & 0 & 0 & 0 & 1 & 1\\1 & 0 & 1 & 0 & 1 & 1 & 1 & 1 & 0 & 1 & 0 & 1 & 1 & 0 & 1 & 0 & 0 & 0 & 0\\1 & 0 & 1 & 1 & 1 & 0 & 0 & 0 & 0 & 0 & 1 & 1 & 0 & 0 & 0 & 1 & 0 & 1 & 1\\1 & 0 & 1 & 1 & 0 & 0 & 0 & 0 & 1 & 0 & 0 & 1 & 1 & 0 & 0 & 1 & 1 & 1 & 0\\0 & 1 & 1 & 0 & 0 & 1 & 1 & 1 & 0 & 1 & 0 & 0 & 0 & 0 & 1 & 1 & 1 & 0 & 0\\1 & 1 & 0 & 0 & 1 & 1 & 0 & 1 & 0 & 0 & 1 & 1 & 1 & 1 & 0 & 0 & 0 & 0 & 1\\1 & 1 & 0 & 0 & 1 & 0 & 1 & 0 & 1 & 1 & 0 & 1 & 0 & 0 & 0 & 0 & 1 & 1 & 0\\1 & 0 & 1 & 0 & 1 & 0 & 0 & 1 & 1 & 0 & 1 & 1 & 1 & 0 & 0 & 1 & 0 & 1 & 0\\0 & 1 & 1 & 0 & 1 & 1 & 0 & 1 & 0 & 0 & 1 & 1 & 0 & 0 & 1 & 0 & 0 & 1 & 0\\0 & 0 & 1 & 1 & 0 & 1 & 1 & 0 & 1 & 1 & 1 & 1 & 0 & 0 & 1 & 0 & 0 & 0 & 1\\1 & 1 & 0 & 0 & 0 & 0 & 0 & 1 & 1 & 0 & 1 & 0 & 1 & 0 & 1 & 0 & 1 & 1 & 1\\0 & 1 & 1 & 0 & 0 & 1 & 1 & 1 & 0 & 1 & 1 & 0 & 0 & 0 & 0 & 1 & 0 & 1 & 1\\0 & 1 & 0 & 0 & 1 & 0 & 1 & 1 & 1 & 0 & 0 & 1 & 1 & 0 & 0 & 1 & 0 & 0 & 1\\0 & 0 & 0 & 1 & 0 & 1 & 1 & 0 & 0 & 1 & 1 & 1 & 0 & 1 & 0 & 0 & 1 & 1 & 1\\1 & 0 & 1 & 1 & 0 & 0 & 0 & 0 & 1 & 1 & 1 & 0 & 1 & 1 & 1 & 0 & 0 & 0 & 1\\1 & 1 & 1 & 0 & 0 & 1 & 1 & 0 & 1 & 0 & 0 & 0 & 0 & 1 & 0 & 1 & 1 & 0 & 1\\0 & 0 & 0 & 0 & 1 & 1 & 1 & 0 & 1 & 0 & 0 & 0 & 0 & 1 & 1 & 1 & 0 & 1 & 1\\1 & 0 & 1 & 0 & 1 & 1 & 0 & 1 & 0 & 1 & 0 & 0 & 0 & 0 & 1 & 0 & 1 & 0 & 1\\1 & 1 & 1 & 1 & 1 & 0 & 0 & 1 & 0 & 0 & 0 & 1 & 0 & 0 & 0 & 1 & 1 & 1 & 0\end{array}\right]}\\
	\mathbf{A}_{m=21}=\scalemath{0.5}{\left[\begin{array}{ccccccccccccccccccccc}0 & 0 & 0 & 0 & 0 & 0 & 0 & 0 & 0 & 0 & 1 & 1 & 1 & 1 & 1 & 1 & 1 & 1 & 1 & 1 & 1\\0 & 1 & 1 & 0 & 1 & 1 & 1 & 0 & 0 & 1 & 1 & 0 & 0 & 0 & 1 & 0 & 0 & 0 & 1 & 1 & 1\\0 & 1 & 0 & 0 & 1 & 0 & 1 & 1 & 0 & 1 & 1 & 0 & 0 & 0 & 1 & 1 & 1 & 0 & 1 & 1 & 0\\0 & 1 & 0 & 0 & 1 & 0 & 0 & 1 & 0 & 1 & 0 & 1 & 0 & 1 & 0 & 1 & 1 & 0 & 1 & 1 & 0\\0 & 1 & 0 & 1 & 1 & 1 & 0 & 0 & 0 & 1 & 1 & 0 & 1 & 0 & 1 & 1 & 0 & 0 & 0 & 1 & 1\\1 & 0 & 0 & 0 & 1 & 0 & 1 & 0 & 1 & 0 & 0 & 0 & 0 & 1 & 0 & 1 & 1 & 1 & 1 & 1 & 0\\1 & 1 & 0 & 0 & 0 & 1 & 0 & 1 & 0 & 1 & 1 & 1 & 0 & 0 & 0 & 1 & 0 & 0 & 0 & 1 & 1\\0 & 0 & 1 & 0 & 0 & 0 & 0 & 0 & 1 & 0 & 1 & 1 & 1 & 1 & 1 & 1 & 1 & 0 & 0 & 1 & 1\\1 & 0 & 1 & 0 & 0 & 0 & 0 & 0 & 1 & 0 & 1 & 1 & 0 & 1 & 1 & 0 & 1 & 1 & 0 & 0 & 1\\0 & 1 & 0 & 0 & 1 & 1 & 0 & 1 & 0 & 0 & 1 & 1 & 0 & 1 & 1 & 1 & 0 & 1 & 0 & 1 & 0\\1 & 0 & 1 & 1 & 0 & 1 & 0 & 0 & 1 & 0 & 0 & 1 & 0 & 1 & 1 & 0 & 1 & 0 & 1 & 0 & 1\\1 & 1 & 1 & 0 & 1 & 0 & 1 & 0 & 1 & 1 & 1 & 0 & 0 & 0 & 0 & 1 & 0 & 1 & 0 & 1 & 0\\0 & 0 & 1 & 1 & 0 & 0 & 0 & 1 & 1 & 1 & 0 & 1 & 1 & 1 & 0 & 1 & 1 & 0 & 0 & 0 & 1\\0 & 1 & 0 & 0 & 0 & 1 & 1 & 1 & 1 & 0 & 0 & 1 & 1 & 1 & 1 & 1 & 0 & 1 & 0 & 0 & 0\\1 & 0 & 0 & 0 & 1 & 0 & 0 & 1 & 0 & 0 & 1 & 1 & 1 & 1 & 1 & 0 & 1 & 1 & 0 & 0 & 0\\0 & 0 & 1 & 1 & 0 & 0 & 1 & 0 & 0 & 0 & 1 & 0 & 1 & 1 & 1 & 1 & 0 & 0 & 1 & 1 & 0\\1 & 1 & 1 & 0 & 1 & 1 & 1 & 0 & 0 & 1 & 0 & 0 & 0 & 0 & 1 & 1 & 0 & 0 & 1 & 0 & 1\\1 & 0 & 0 & 1 & 1 & 0 & 0 & 0 & 1 & 1 & 1 & 1 & 1 & 0 & 0 & 1 & 1 & 0 & 0 & 0 & 1\\0 & 1 & 1 & 0 & 1 & 1 & 0 & 0 & 1 & 0 & 1 & 0 & 0 & 0 & 1 & 0 & 1 & 0 & 1 & 0 & 1\\0 & 0 & 0 & 1 & 1 & 0 & 0 & 0 & 0 & 1 & 1 & 1 & 0 & 0 & 1 & 1 & 0 & 1 & 1 & 1 & 0\\1 & 0 & 0 & 0 & 0 & 0 & 1 & 1 & 1 & 1 & 0 & 0 & 1 & 1 & 0 & 0 & 1 & 0 & 1 & 1 & 0\\0 & 1 & 1 & 0 & 0 & 1 & 1 & 1 & 1 & 0 & 0 & 0 & 0 & 1 & 0 & 1 & 1 & 1 & 0 & 1 & 0\end{array}\right]}\\
	\mathbf{A}_{m=23}=\scalemath{0.5}{\left[\begin{array}{ccccccccccccccccccccccc}0 & 0 & 0 & 0 & 0 & 0 & 0 & 0 & 0 & 0 & 0 & 1 & 1 & 1 & 1 & 1 & 1 & 1 & 1 & 1 & 1 & 1 & 1\\1 & 0 & 0 & 1 & 0 & 0 & 1 & 1 & 1 & 0 & 0 & 1 & 1 & 0 & 1 & 0 & 0 & 1 & 0 & 1 & 1 & 0 & 1\\1 & 0 & 0 & 1 & 1 & 0 & 1 & 1 & 0 & 0 & 1 & 1 & 1 & 1 & 0 & 1 & 0 & 1 & 0 & 0 & 0 & 1 & 0\\0 & 1 & 0 & 0 & 1 & 1 & 1 & 1 & 0 & 1 & 1 & 1 & 0 & 1 & 1 & 1 & 0 & 0 & 0 & 1 & 0 & 0 & 0\\0 & 0 & 0 & 1 & 1 & 1 & 1 & 0 & 1 & 0 & 1 & 1 & 0 & 0 & 0 & 0 & 1 & 0 & 1 & 1 & 0 & 0 & 1\\1 & 1 & 0 & 0 & 1 & 0 & 0 & 0 & 0 & 0 & 1 & 0 & 0 & 1 & 1 & 1 & 1 & 1 & 0 & 0 & 1 & 0 & 1\\1 & 1 & 0 & 0 & 0 & 1 & 1 & 1 & 1 & 1 & 1 & 0 & 0 & 0 & 0 & 1 & 1 & 1 & 0 & 0 & 0 & 1 & 0\\0 & 0 & 1 & 0 & 0 & 0 & 0 & 1 & 1 & 1 & 1 & 0 & 0 & 0 & 0 & 1 & 1 & 1 & 1 & 1 & 0 & 1 & 0\\0 & 0 & 0 & 1 & 0 & 1 & 1 & 0 & 1 & 1 & 1 & 1 & 1 & 0 & 1 & 0 & 0 & 1 & 1 & 0 & 1 & 0 & 0\\0 & 0 & 1 & 0 & 1 & 1 & 0 & 1 & 1 & 0 & 0 & 1 & 0 & 0 & 1 & 0 & 0 & 1 & 0 & 0 & 1 & 1 & 1\\1 & 0 & 0 & 0 & 0 & 0 & 0 & 0 & 1 & 1 & 0 & 0 & 1 & 1 & 1 & 0 & 1 & 0 & 1 & 1 & 1 & 1 & 1\\0 & 0 & 1 & 1 & 1 & 1 & 0 & 0 & 1 & 1 & 1 & 0 & 1 & 0 & 0 & 0 & 1 & 1 & 0 & 1 & 0 & 0 & 1\\1 & 1 & 0 & 0 & 1 & 1 & 0 & 1 & 0 & 0 & 0 & 0 & 1 & 0 & 0 & 1 & 1 & 0 & 1 & 0 & 1 & 0 & 1\\0 & 1 & 0 & 0 & 1 & 1 & 1 & 1 & 1 & 0 & 1 & 1 & 0 & 0 & 0 & 0 & 0 & 1 & 1 & 1 & 1 & 0 & 0\\0 & 1 & 1 & 1 & 1 & 0 & 0 & 0 & 0 & 1 & 0 & 1 & 1 & 0 & 0 & 1 & 0 & 1 & 0 & 0 & 1 & 1 & 1\\0 & 1 & 1 & 1 & 1 & 0 & 1 & 1 & 1 & 0 & 0 & 0 & 1 & 0 & 0 & 0 & 1 & 1 & 0 & 1 & 0 & 0 & 0\\1 & 0 & 0 & 1 & 0 & 1 & 1 & 0 & 1 & 0 & 0 & 1 & 0 & 1 & 0 & 0 & 0 & 1 & 0 & 1 & 1 & 0 & 1\\1 & 0 & 0 & 1 & 0 & 0 & 1 & 1 & 0 & 0 & 0 & 1 & 0 & 0 & 1 & 0 & 1 & 1 & 1 & 1 & 0 & 1 & 1\\0 & 1 & 1 & 1 & 0 & 1 & 1 & 1 & 0 & 0 & 1 & 1 & 0 & 0 & 0 & 1 & 1 & 0 & 0 & 1 & 0 & 0 & 1\\1 & 0 & 0 & 1 & 1 & 0 & 0 & 1 & 0 & 0 & 1 & 0 & 0 & 1 & 1 & 1 & 0 & 1 & 1 & 0 & 1 & 1 & 0\\1 & 1 & 0 & 0 & 1 & 1 & 0 & 1 & 0 & 0 & 1 & 0 & 1 & 0 & 0 & 0 & 1 & 1 & 0 & 1 & 0 & 1 & 0\\0 & 1 & 0 & 0 & 1 & 0 & 1 & 1 & 0 & 0 & 1 & 0 & 0 & 0 & 0 & 0 & 1 & 1 & 1 & 1 & 1 & 1 & 1\\0 & 1 & 1 & 0 & 0 & 0 & 0 & 1 & 1 & 1 & 1 & 1 & 1 & 0 & 1 & 0 & 0 & 1 & 1 & 1 & 0 & 0 & 0\\1 & 0 & 1 & 1 & 0 & 1 & 1 & 1 & 1 & 1 & 1 & 0 & 1 & 0 & 0 & 1 & 0 & 0 & 0 & 0 & 1 & 0 & 0\end{array}\right]}\\
	\mathbf{A}_{m=25}=\scalemath{0.5}{\left[\begin{array}{ccccccccccccccccccccccccc}0 & 0 & 0 & 0 & 0 & 0 & 0 & 0 & 0 & 0 & 0 & 0 & 1 & 1 & 1 & 1 & 1 & 1 & 1 & 1 & 1 & 1 & 1 & 1 & 1\\0 & 0 & 1 & 0 & 1 & 1 & 1 & 0 & 1 & 0 & 0 & 1 & 0 & 1 & 0 & 0 & 0 & 1 & 0 & 0 & 1 & 1 & 1 & 1 & 1\\1 & 0 & 1 & 1 & 1 & 0 & 0 & 0 & 1 & 0 & 0 & 0 & 0 & 1 & 0 & 0 & 0 & 0 & 1 & 1 & 1 & 0 & 1 & 1 & 1\\1 & 0 & 1 & 1 & 0 & 1 & 0 & 0 & 1 & 0 & 1 & 1 & 0 & 0 & 0 & 1 & 0 & 0 & 1 & 0 & 1 & 1 & 0 & 0 & 1\\0 & 1 & 0 & 1 & 0 & 0 & 1 & 0 & 1 & 1 & 0 & 1 & 1 & 0 & 1 & 1 & 0 & 1 & 0 & 0 & 1 & 1 & 1 & 0 & 0\\1 & 0 & 1 & 0 & 1 & 0 & 0 & 0 & 0 & 1 & 1 & 1 & 1 & 0 & 1 & 0 & 0 & 1 & 0 & 0 & 1 & 1 & 0 & 1 & 0\\1 & 1 & 1 & 1 & 0 & 0 & 1 & 1 & 1 & 0 & 1 & 0 & 1 & 0 & 0 & 0 & 1 & 1 & 0 & 1 & 0 & 1 & 0 & 0 & 0\\1 & 0 & 1 & 1 & 1 & 1 & 1 & 0 & 0 & 0 & 1 & 0 & 0 & 0 & 1 & 1 & 0 & 0 & 1 & 0 & 0 & 1 & 0 & 1 & 1\\1 & 1 & 1 & 1 & 0 & 0 & 0 & 1 & 1 & 1 & 1 & 0 & 1 & 0 & 0 & 0 & 0 & 0 & 1 & 1 & 1 & 0 & 0 & 0 & 0\\1 & 0 & 0 & 0 & 1 & 0 & 1 & 1 & 1 & 1 & 0 & 1 & 0 & 0 & 0 & 0 & 1 & 0 & 0 & 1 & 0 & 0 & 1 & 1 & 1\\0 & 1 & 0 & 0 & 1 & 1 & 0 & 1 & 0 & 0 & 0 & 1 & 1 & 1 & 0 & 0 & 1 & 1 & 1 & 1 & 0 & 1 & 0 & 0 & 0\\0 & 0 & 0 & 1 & 1 & 0 & 1 & 1 & 0 & 1 & 1 & 1 & 0 & 1 & 0 & 1 & 0 & 1 & 0 & 1 & 1 & 0 & 1 & 0 & 0\\0 & 1 & 1 & 1 & 0 & 0 & 1 & 1 & 0 & 0 & 0 & 1 & 1 & 1 & 1 & 1 & 1 & 0 & 1 & 0 & 0 & 1 & 0 & 0 & 0\\1 & 1 & 1 & 0 & 0 & 1 & 1 & 1 & 1 & 0 & 0 & 0 & 0 & 0 & 0 & 1 & 1 & 0 & 0 & 1 & 0 & 0 & 1 & 1 & 0\\0 & 0 & 1 & 0 & 0 & 0 & 1 & 1 & 0 & 1 & 1 & 1 & 1 & 1 & 1 & 1 & 0 & 0 & 0 & 1 & 0 & 0 & 1 & 1 & 0\\0 & 0 & 1 & 1 & 0 & 1 & 0 & 1 & 0 & 0 & 0 & 0 & 1 & 0 & 1 & 1 & 1 & 1 & 0 & 1 & 0 & 0 & 0 & 1 & 1\\1 & 0 & 0 & 1 & 1 & 1 & 0 & 0 & 1 & 0 & 0 & 1 & 0 & 1 & 1 & 0 & 0 & 1 & 1 & 0 & 1 & 0 & 0 & 1 & 1\\0 & 0 & 0 & 0 & 0 & 1 & 1 & 1 & 0 & 0 & 1 & 0 & 0 & 1 & 1 & 0 & 1 & 0 & 1 & 0 & 1 & 1 & 1 & 0 & 1\\0 & 1 & 1 & 0 & 1 & 1 & 1 & 0 & 1 & 1 & 0 & 0 & 0 & 1 & 1 & 1 & 0 & 0 & 0 & 1 & 0 & 0 & 0 & 1 & 0\\1 & 0 & 1 & 1 & 0 & 1 & 1 & 0 & 1 & 1 & 0 & 0 & 1 & 1 & 0 & 0 & 0 & 0 & 0 & 0 & 1 & 0 & 0 & 1 & 1\\0 & 1 & 0 & 0 & 0 & 0 & 1 & 0 & 0 & 1 & 0 & 1 & 0 & 1 & 1 & 0 & 1 & 1 & 0 & 0 & 1 & 0 & 1 & 1 & 1\\0 & 0 & 0 & 1 & 0 & 1 & 1 & 1 & 1 & 0 & 0 & 0 & 1 & 0 & 1 & 1 & 0 & 0 & 0 & 1 & 1 & 1 & 1 & 1 & 0\\1 & 1 & 1 & 0 & 1 & 0 & 0 & 0 & 0 & 1 & 1 & 0 & 0 & 1 & 0 & 0 & 0 & 1 & 1 & 1 & 1 & 0 & 0 & 1 & 0\\1 & 0 & 0 & 0 & 1 & 1 & 0 & 1 & 0 & 0 & 0 & 0 & 1 & 0 & 1 & 1 & 1 & 0 & 1 & 1 & 1 & 1 & 0 & 1 & 0\\1 & 1 & 0 & 0 & 1 & 0 & 1 & 0 & 1 & 0 & 0 & 0 & 1 & 1 & 1 & 0 & 1 & 1 & 1 & 0 & 0 & 0 & 1 & 0 & 0\\1 & 1 & 0 & 1 & 1 & 0 & 0 & 1 & 0 & 0 & 1 & 1 & 1 & 0 & 0 & 1 & 0 & 1 & 0 & 0 & 1 & 0 & 1 & 0 & 0\end{array}\right]}\\
	\mathbf{A}_{m=27}=\scalemath{0.5}{\left[\begin{array}{ccccccccccccccccccccccccccc}0 & 0 & 0 & 0 & 0 & 0 & 0 & 0 & 0 & 0 & 0 & 0 & 0 & 1 & 1 & 1 & 1 & 1 & 1 & 1 & 1 & 1 & 1 & 1 & 1 & 1 & 1\\0 & 1 & 1 & 0 & 1 & 1 & 0 & 0 & 0 & 1 & 1 & 0 & 0 & 1 & 0 & 0 & 1 & 0 & 1 & 1 & 1 & 0 & 1 & 0 & 1 & 1 & 0\\0 & 1 & 0 & 0 & 0 & 0 & 0 & 0 & 1 & 1 & 1 & 0 & 1 & 0 & 1 & 0 & 1 & 1 & 1 & 1 & 1 & 0 & 0 & 1 & 0 & 1 & 0\\1 & 1 & 0 & 1 & 0 & 1 & 1 & 1 & 1 & 0 & 0 & 1 & 1 & 0 & 0 & 1 & 0 & 0 & 0 & 1 & 0 & 1 & 0 & 0 & 0 & 1 & 0\\0 & 0 & 1 & 0 & 1 & 0 & 1 & 1 & 0 & 1 & 1 & 1 & 0 & 1 & 1 & 0 & 0 & 0 & 0 & 0 & 0 & 0 & 1 & 1 & 0 & 1 & 1\\1 & 0 & 0 & 0 & 0 & 1 & 1 & 1 & 0 & 0 & 1 & 0 & 1 & 1 & 1 & 0 & 1 & 1 & 0 & 1 & 0 & 1 & 0 & 1 & 1 & 0 & 0\\0 & 0 & 0 & 0 & 1 & 0 & 1 & 1 & 1 & 1 & 1 & 0 & 1 & 0 & 0 & 1 & 1 & 0 & 1 & 0 & 1 & 1 & 0 & 0 & 0 & 0 & 1\\0 & 0 & 1 & 1 & 0 & 1 & 1 & 0 & 1 & 0 & 0 & 0 & 1 & 1 & 0 & 1 & 1 & 1 & 1 & 0 & 1 & 0 & 0 & 1 & 0 & 0 & 0\\1 & 1 & 1 & 1 & 0 & 1 & 0 & 1 & 1 & 0 & 0 & 0 & 0 & 0 & 1 & 1 & 1 & 0 & 0 & 0 & 1 & 0 & 0 & 1 & 0 & 0 & 1\\1 & 0 & 0 & 1 & 1 & 1 & 0 & 1 & 1 & 1 & 1 & 0 & 1 & 0 & 1 & 0 & 1 & 1 & 0 & 1 & 0 & 0 & 0 & 0 & 0 & 1 & 0\\0 & 1 & 1 & 0 & 1 & 0 & 1 & 1 & 1 & 1 & 0 & 0 & 1 & 0 & 0 & 0 & 1 & 0 & 0 & 1 & 1 & 0 & 1 & 1 & 0 & 0 & 0\\0 & 0 & 0 & 0 & 1 & 1 & 0 & 1 & 1 & 0 & 0 & 0 & 1 & 1 & 1 & 1 & 0 & 1 & 0 & 0 & 1 & 1 & 0 & 1 & 0 & 1 & 1\\1 & 1 & 1 & 0 & 0 & 1 & 1 & 0 & 1 & 1 & 1 & 1 & 0 & 0 & 1 & 1 & 0 & 0 & 1 & 0 & 0 & 0 & 0 & 1 & 1 & 0 & 0\\1 & 0 & 1 & 0 & 0 & 0 & 0 & 1 & 0 & 1 & 1 & 1 & 0 & 1 & 1 & 0 & 1 & 1 & 0 & 0 & 1 & 0 & 1 & 0 & 1 & 0 & 0\\0 & 1 & 1 & 0 & 0 & 1 & 0 & 0 & 1 & 1 & 0 & 0 & 0 & 1 & 0 & 1 & 0 & 1 & 1 & 0 & 1 & 1 & 0 & 0 & 1 & 0 & 1\\1 & 0 & 1 & 0 & 1 & 0 & 0 & 1 & 1 & 0 & 1 & 0 & 0 & 1 & 1 & 0 & 1 & 0 & 1 & 0 & 0 & 1 & 1 & 0 & 0 & 0 & 1\\0 & 1 & 1 & 0 & 1 & 0 & 1 & 0 & 0 & 0 & 1 & 0 & 0 & 1 & 0 & 1 & 0 & 1 & 0 & 1 & 0 & 0 & 1 & 1 & 1 & 1 & 1\\1 & 1 & 0 & 0 & 0 & 0 & 0 & 1 & 1 & 0 & 0 & 1 & 0 & 1 & 0 & 1 & 0 & 1 & 1 & 0 & 1 & 1 & 0 & 1 & 1 & 1 & 0\\1 & 0 & 0 & 1 & 1 & 1 & 0 & 1 & 0 & 0 & 0 & 0 & 1 & 0 & 0 & 1 & 1 & 1 & 1 & 1 & 0 & 1 & 0 & 0 & 1 & 0 & 0\\0 & 0 & 0 & 0 & 0 & 0 & 0 & 0 & 1 & 0 & 0 & 1 & 1 & 0 & 1 & 1 & 1 & 1 & 1 & 0 & 1 & 1 & 1 & 1 & 1 & 1 & 0\\0 & 0 & 0 & 1 & 1 & 0 & 0 & 1 & 1 & 1 & 1 & 0 & 0 & 1 & 1 & 1 & 0 & 1 & 1 & 0 & 0 & 0 & 0 & 0 & 0 & 1 & 1\\0 & 1 & 0 & 0 & 1 & 1 & 0 & 1 & 1 & 0 & 0 & 1 & 1 & 0 & 0 & 1 & 0 & 1 & 0 & 0 & 1 & 1 & 0 & 1 & 0 & 1 & 0\\1 & 0 & 1 & 1 & 1 & 0 & 0 & 1 & 1 & 0 & 1 & 1 & 0 & 1 & 0 & 1 & 1 & 0 & 0 & 0 & 0 & 1 & 0 & 1 & 0 & 1 & 0\\0 & 0 & 1 & 0 & 0 & 0 & 1 & 1 & 0 & 1 & 1 & 1 & 1 & 0 & 1 & 0 & 0 & 1 & 1 & 1 & 0 & 1 & 1 & 0 & 0 & 1 & 0\\1 & 1 & 0 & 1 & 0 & 0 & 1 & 1 & 0 & 0 & 1 & 1 & 1 & 0 & 0 & 0 & 1 & 0 & 0 & 1 & 1 & 0 & 0 & 1 & 0 & 1 & 0\\1 & 1 & 0 & 0 & 0 & 0 & 0 & 0 & 0 & 1 & 0 & 1 & 1 & 0 & 1 & 1 & 0 & 0 & 1 & 1 & 1 & 0 & 0 & 1 & 1 & 1 & 1\\1 & 0 & 1 & 1 & 0 & 1 & 0 & 1 & 0 & 1 & 1 & 0 & 0 & 1 & 1 & 0 & 0 & 0 & 0 & 0 & 1 & 1 & 1 & 0 & 1 & 0 & 0\\1 & 0 & 1 & 1 & 0 & 0 & 1 & 0 & 0 & 1 & 1 & 1 & 1 & 0 & 0 & 1 & 0 & 1 & 0 & 1 & 0 & 1 & 0 & 1 & 1 & 0 & 0\end{array}\right]}\\
	\mathbf{A}_{m=29}=\scalemath{0.5}{\left[\begin{array}{ccccccccccccccccccccccccccccc}0 & 0 & 0 & 0 & 0 & 0 & 0 & 0 & 0 & 0 & 0 & 0 & 0 & 0 & 1 & 1 & 1 & 1 & 1 & 1 & 1 & 1 & 1 & 1 & 1 & 1 & 1 & 1 & 1\\1 & 0 & 0 & 1 & 1 & 0 & 1 & 1 & 0 & 1 & 1 & 1 & 1 & 0 & 1 & 0 & 0 & 1 & 0 & 0 & 0 & 0 & 0 & 1 & 0 & 1 & 1 & 0 & 1\\1 & 0 & 0 & 1 & 0 & 1 & 0 & 1 & 1 & 0 & 1 & 0 & 1 & 0 & 0 & 0 & 0 & 1 & 0 & 0 & 1 & 1 & 1 & 1 & 1 & 1 & 0 & 0 & 0\\0 & 1 & 1 & 1 & 0 & 1 & 0 & 1 & 1 & 0 & 0 & 0 & 1 & 1 & 1 & 0 & 0 & 0 & 0 & 0 & 1 & 1 & 1 & 0 & 0 & 0 & 1 & 1 & 1\\0 & 0 & 1 & 0 & 1 & 0 & 1 & 0 & 1 & 0 & 1 & 0 & 1 & 1 & 1 & 1 & 1 & 0 & 1 & 0 & 1 & 0 & 0 & 1 & 0 & 0 & 0 & 1 & 1\\0 & 1 & 1 & 0 & 0 & 0 & 1 & 0 & 1 & 1 & 1 & 1 & 1 & 1 & 0 & 1 & 0 & 1 & 0 & 0 & 0 & 0 & 0 & 1 & 1 & 1 & 0 & 0 & 1\\1 & 1 & 0 & 0 & 0 & 0 & 0 & 0 & 0 & 1 & 0 & 1 & 0 & 1 & 1 & 1 & 1 & 1 & 1 & 1 & 1 & 0 & 0 & 0 & 0 & 1 & 0 & 1 & 1\\0 & 0 & 0 & 0 & 0 & 0 & 1 & 1 & 1 & 1 & 0 & 1 & 0 & 1 & 0 & 1 & 1 & 0 & 0 & 1 & 1 & 0 & 1 & 1 & 0 & 0 & 0 & 1 & 1\\1 & 1 & 1 & 0 & 1 & 0 & 1 & 1 & 1 & 0 & 1 & 1 & 1 & 0 & 0 & 1 & 0 & 0 & 0 & 0 & 0 & 0 & 0 & 1 & 0 & 1 & 1 & 0 & 0\\1 & 1 & 0 & 1 & 1 & 0 & 0 & 1 & 1 & 0 & 0 & 1 & 1 & 0 & 0 & 0 & 1 & 1 & 1 & 0 & 1 & 0 & 0 & 0 & 0 & 1 & 1 & 0 & 1\\0 & 0 & 0 & 1 & 1 & 1 & 0 & 0 & 1 & 1 & 1 & 1 & 0 & 1 & 0 & 0 & 0 & 1 & 1 & 1 & 0 & 0 & 1 & 1 & 0 & 0 & 1 & 0 & 0\\0 & 0 & 0 & 0 & 0 & 1 & 0 & 1 & 1 & 1 & 0 & 1 & 0 & 1 & 1 & 0 & 1 & 1 & 0 & 0 & 1 & 1 & 1 & 0 & 0 & 1 & 1 & 0 & 1\\1 & 0 & 1 & 1 & 0 & 0 & 0 & 0 & 0 & 1 & 0 & 0 & 1 & 0 & 1 & 1 & 1 & 1 & 1 & 1 & 0 & 1 & 0 & 0 & 0 & 0 & 1 & 1 & 0\\1 & 1 & 1 & 1 & 1 & 0 & 1 & 0 & 0 & 1 & 0 & 1 & 0 & 1 & 1 & 1 & 0 & 0 & 1 & 0 & 1 & 1 & 0 & 0 & 0 & 0 & 1 & 0 & 0\\1 & 1 & 1 & 1 & 0 & 1 & 0 & 1 & 0 & 0 & 1 & 1 & 0 & 0 & 0 & 1 & 0 & 0 & 1 & 0 & 1 & 1 & 0 & 1 & 0 & 0 & 1 & 0 & 1\\1 & 0 & 1 & 1 & 1 & 0 & 1 & 0 & 0 & 1 & 0 & 1 & 1 & 1 & 1 & 0 & 1 & 0 & 1 & 0 & 1 & 1 & 1 & 0 & 0 & 0 & 0 & 0 & 0\\1 & 0 & 0 & 1 & 1 & 1 & 0 & 0 & 0 & 0 & 1 & 1 & 1 & 0 & 1 & 0 & 0 & 0 & 0 & 1 & 1 & 0 & 1 & 0 & 1 & 1 & 0 & 0 & 1\\1 & 1 & 0 & 1 & 0 & 0 & 1 & 1 & 1 & 0 & 1 & 0 & 1 & 0 & 0 & 0 & 0 & 0 & 0 & 1 & 1 & 1 & 1 & 1 & 1 & 0 & 0 & 1 & 0\\0 & 1 & 1 & 0 & 1 & 0 & 0 & 1 & 1 & 0 & 0 & 1 & 1 & 1 & 1 & 0 & 0 & 0 & 0 & 1 & 0 & 0 & 1 & 1 & 0 & 1 & 1 & 0 & 1\\1 & 0 & 1 & 0 & 1 & 1 & 0 & 0 & 0 & 0 & 0 & 1 & 0 & 1 & 0 & 1 & 1 & 1 & 0 & 1 & 1 & 0 & 0 & 0 & 1 & 0 & 1 & 1 & 1\\0 & 0 & 1 & 0 & 0 & 1 & 1 & 0 & 1 & 1 & 1 & 1 & 1 & 0 & 0 & 0 & 0 & 0 & 0 & 1 & 1 & 1 & 0 & 0 & 1 & 1 & 0 & 1 & 1\\1 & 0 & 1 & 0 & 1 & 1 & 0 & 1 & 0 & 1 & 1 & 1 & 0 & 0 & 1 & 0 & 0 & 1 & 0 & 0 & 0 & 0 & 0 & 1 & 1 & 1 & 1 & 0 & 1\\0 & 0 & 1 & 0 & 1 & 0 & 0 & 0 & 1 & 0 & 1 & 1 & 1 & 1 & 1 & 1 & 1 & 0 & 0 & 1 & 0 & 0 & 0 & 1 & 1 & 0 & 1 & 0 & 1\\1 & 0 & 0 & 1 & 0 & 1 & 0 & 1 & 0 & 1 & 0 & 1 & 1 & 1 & 1 & 1 & 0 & 1 & 1 & 0 & 0 & 0 & 0 & 0 & 0 & 0 & 1 & 0 & 1\\0 & 0 & 0 & 1 & 1 & 1 & 1 & 0 & 1 & 1 & 0 & 1 & 0 & 0 & 1 & 0 & 1 & 0 & 0 & 1 & 1 & 0 & 1 & 0 & 0 & 1 & 0 & 1 & 1\\0 & 1 & 1 & 0 & 1 & 1 & 0 & 1 & 1 & 1 & 1 & 1 & 0 & 1 & 1 & 0 & 0 & 1 & 1 & 0 & 0 & 0 & 1 & 0 & 0 & 1 & 0 & 0 & 0\\1 & 0 & 0 & 0 & 0 & 1 & 1 & 1 & 1 & 1 & 1 & 1 & 0 & 0 & 1 & 1 & 0 & 1 & 0 & 0 & 1 & 0 & 0 & 1 & 0 & 0 & 1 & 1 & 0\\0 & 1 & 0 & 0 & 1 & 0 & 1 & 1 & 1 & 0 & 0 & 0 & 1 & 0 & 1 & 0 & 1 & 0 & 0 & 1 & 1 & 0 & 1 & 0 & 0 & 1 & 1 & 0 & 1\\0 & 1 & 0 & 0 & 1 & 1 & 0 & 0 & 0 & 0 & 1 & 1 & 1 & 0 & 1 & 1 & 0 & 1 & 0 & 0 & 1 & 1 & 0 & 1 & 0 & 0 & 1 & 1 & 0\\1 & 1 & 1 & 0 & 1 & 1 & 0 & 0 & 1 & 1 & 0 & 0 & 1 & 1 & 0 & 1 & 0 & 1 & 0 & 0 & 1 & 0 & 0 & 1 & 1 & 0 & 0 & 0 & 1\end{array}\right]}\\
	\mathbf{A}_{m=31}=\scalemath{0.5}{\left[\begin{array}{ccccccccccccccccccccccccccccccc}0 & 0 & 0 & 0 & 0 & 0 & 0 & 0 & 0 & 0 & 0 & 0 & 0 & 0 & 0 & 1 & 1 & 1 & 1 & 1 & 1 & 1 & 1 & 1 & 1 & 1 & 1 & 1 & 1 & 1 & 1\\0 & 0 & 0 & 1 & 1 & 0 & 1 & 1 & 1 & 0 & 0 & 0 & 1 & 0 & 1 & 1 & 0 & 0 & 1 & 1 & 0 & 1 & 1 & 0 & 0 & 0 & 0 & 0 & 1 & 1 & 1\\0 & 1 & 1 & 1 & 0 & 1 & 0 & 1 & 1 & 1 & 1 & 1 & 0 & 1 & 0 & 1 & 0 & 0 & 1 & 0 & 0 & 0 & 0 & 0 & 0 & 0 & 1 & 0 & 1 & 1 & 0\\1 & 0 & 0 & 1 & 1 & 1 & 0 & 1 & 1 & 1 & 1 & 0 & 0 & 0 & 1 & 0 & 1 & 0 & 1 & 1 & 1 & 1 & 0 & 0 & 1 & 1 & 0 & 0 & 0 & 0 & 0\\0 & 1 & 0 & 0 & 1 & 0 & 1 & 1 & 0 & 1 & 1 & 1 & 1 & 1 & 1 & 0 & 1 & 0 & 0 & 0 & 0 & 1 & 1 & 1 & 1 & 0 & 0 & 0 & 0 & 0 & 1\\1 & 0 & 0 & 0 & 1 & 1 & 0 & 1 & 1 & 0 & 0 & 1 & 1 & 0 & 0 & 0 & 1 & 1 & 0 & 0 & 1 & 1 & 0 & 1 & 1 & 0 & 1 & 0 & 1 & 1 & 0\\1 & 0 & 1 & 1 & 1 & 0 & 0 & 0 & 0 & 1 & 1 & 1 & 0 & 0 & 0 & 0 & 1 & 1 & 0 & 0 & 1 & 0 & 1 & 0 & 0 & 1 & 0 & 1 & 0 & 1 & 1\\0 & 0 & 1 & 0 & 1 & 0 & 1 & 1 & 0 & 0 & 0 & 0 & 1 & 1 & 0 & 1 & 0 & 0 & 0 & 1 & 1 & 1 & 0 & 0 & 0 & 1 & 1 & 1 & 1 & 1 & 0\\0 & 0 & 1 & 0 & 0 & 1 & 1 & 0 & 0 & 0 & 0 & 1 & 0 & 0 & 0 & 1 & 0 & 1 & 1 & 0 & 1 & 1 & 1 & 0 & 1 & 1 & 0 & 1 & 1 & 1 & 1\\0 & 0 & 0 & 1 & 1 & 1 & 0 & 1 & 1 & 0 & 0 & 1 & 0 & 0 & 1 & 1 & 1 & 1 & 0 & 0 & 1 & 0 & 1 & 0 & 0 & 0 & 0 & 1 & 1 & 1 & 0\\0 & 0 & 0 & 1 & 0 & 0 & 1 & 1 & 1 & 1 & 1 & 0 & 1 & 1 & 1 & 0 & 1 & 0 & 1 & 0 & 1 & 0 & 1 & 1 & 1 & 0 & 0 & 0 & 0 & 0 & 1\\0 & 1 & 1 & 1 & 0 & 0 & 1 & 0 & 1 & 0 & 0 & 1 & 1 & 0 & 0 & 0 & 0 & 0 & 0 & 1 & 1 & 1 & 0 & 1 & 1 & 1 & 1 & 0 & 0 & 0 & 1\\0 & 1 & 0 & 0 & 0 & 1 & 0 & 1 & 1 & 0 & 1 & 1 & 0 & 1 & 1 & 0 & 1 & 1 & 1 & 0 & 0 & 1 & 1 & 0 & 0 & 0 & 0 & 1 & 0 & 1 & 1\\0 & 1 & 0 & 0 & 1 & 0 & 1 & 1 & 1 & 0 & 0 & 1 & 0 & 1 & 1 & 0 & 0 & 0 & 0 & 0 & 1 & 1 & 1 & 0 & 1 & 1 & 0 & 1 & 0 & 0 & 1\\1 & 0 & 0 & 1 & 0 & 0 & 0 & 1 & 0 & 1 & 0 & 0 & 1 & 1 & 0 & 1 & 1 & 1 & 1 & 1 & 1 & 0 & 0 & 1 & 0 & 1 & 0 & 1 & 0 & 1 & 0\\1 & 1 & 0 & 1 & 0 & 1 & 0 & 0 & 0 & 0 & 1 & 1 & 1 & 0 & 1 & 1 & 0 & 1 & 0 & 1 & 1 & 0 & 0 & 1 & 0 & 0 & 0 & 1 & 1 & 0 & 0\\1 & 1 & 0 & 1 & 1 & 1 & 0 & 1 & 0 & 0 & 0 & 1 & 0 & 1 & 1 & 0 & 0 & 1 & 0 & 1 & 0 & 1 & 1 & 1 & 1 & 0 & 0 & 0 & 0 & 1 & 0\\0 & 0 & 0 & 0 & 0 & 1 & 0 & 0 & 1 & 0 & 1 & 0 & 0 & 1 & 1 & 0 & 0 & 1 & 1 & 1 & 1 & 1 & 0 & 1 & 1 & 1 & 0 & 1 & 0 & 0 & 1\\0 & 0 & 1 & 0 & 0 & 0 & 1 & 1 & 0 & 1 & 1 & 1 & 1 & 0 & 1 & 0 & 1 & 1 & 0 & 1 & 0 & 0 & 0 & 0 & 1 & 0 & 0 & 1 & 1 & 1 & 0\\0 & 0 & 1 & 1 & 1 & 0 & 1 & 0 & 0 & 0 & 0 & 1 & 1 & 0 & 1 & 0 & 0 & 0 & 0 & 1 & 1 & 1 & 0 & 1 & 0 & 1 & 0 & 1 & 1 & 1 & 0\\0 & 0 & 0 & 0 & 0 & 1 & 1 & 0 & 1 & 1 & 1 & 1 & 0 & 0 & 0 & 0 & 0 & 0 & 1 & 0 & 1 & 1 & 0 & 1 & 1 & 0 & 1 & 1 & 1 & 1 & 0\\0 & 0 & 0 & 1 & 0 & 1 & 1 & 1 & 0 & 0 & 0 & 0 & 0 & 0 & 1 & 1 & 1 & 1 & 0 & 0 & 0 & 1 & 1 & 1 & 1 & 0 & 1 & 0 & 1 & 1 & 1\\1 & 0 & 1 & 1 & 1 & 0 & 1 & 1 & 1 & 0 & 0 & 0 & 1 & 0 & 1 & 1 & 1 & 1 & 1 & 1 & 0 & 0 & 0 & 0 & 1 & 0 & 0 & 1 & 0 & 0 & 0\\0 & 0 & 1 & 0 & 0 & 1 & 1 & 1 & 0 & 1 & 0 & 1 & 1 & 1 & 0 & 1 & 0 & 1 & 0 & 1 & 0 & 1 & 1 & 0 & 0 & 0 & 1 & 0 & 0 & 1 & 1\\1 & 0 & 0 & 1 & 0 & 1 & 1 & 0 & 1 & 0 & 0 & 0 & 0 & 0 & 0 & 1 & 0 & 0 & 1 & 1 & 1 & 1 & 0 & 1 & 1 & 0 & 1 & 0 & 1 & 1 & 0\\0 & 0 & 1 & 1 & 1 & 1 & 1 & 0 & 0 & 0 & 0 & 1 & 0 & 1 & 1 & 0 & 0 & 0 & 1 & 0 & 1 & 1 & 1 & 1 & 0 & 0 & 1 & 0 & 1 & 1 & 0\\0 & 1 & 1 & 1 & 0 & 1 & 1 & 1 & 0 & 0 & 0 & 1 & 0 & 0 & 0 & 1 & 1 & 0 & 1 & 0 & 1 & 0 & 0 & 1 & 1 & 0 & 1 & 0 & 1 & 0 & 1\\1 & 0 & 1 & 1 & 0 & 1 & 1 & 0 & 0 & 1 & 0 & 0 & 1 & 1 & 0 & 1 & 1 & 1 & 0 & 1 & 0 & 0 & 0 & 0 & 1 & 1 & 0 & 0 & 0 & 1 & 0\\0 & 0 & 1 & 0 & 0 & 0 & 1 & 0 & 1 & 1 & 1 & 1 & 0 & 1 & 1 & 1 & 1 & 1 & 0 & 1 & 1 & 1 & 0 & 0 & 0 & 0 & 1 & 1 & 0 & 0 & 0\\0 & 0 & 0 & 1 & 0 & 1 & 1 & 1 & 0 & 1 & 0 & 0 & 1 & 1 & 0 & 1 & 0 & 1 & 1 & 1 & 0 & 1 & 0 & 1 & 1 & 0 & 1 & 0 & 1 & 0 & 0\\1 & 0 & 0 & 0 & 0 & 0 & 1 & 0 & 1 & 1 & 0 & 0 & 1 & 1 & 0 & 1 & 1 & 1 & 1 & 0 & 1 & 0 & 0 & 1 & 1 & 0 & 0 & 1 & 1 & 1 & 0\\0 & 0 & 1 & 0 & 1 & 1 & 1 & 0 & 1 & 0 & 0 & 0 & 1 & 0 & 1 & 1 & 0 & 1 & 1 & 1 & 0 & 0 & 0 & 1 & 1 & 1 & 0 & 0 & 1 & 0 & 1\end{array}\right]}
\end{eqnarray*}

\begin{backmatter}

\bmsection{Funding}
National Science Center, Poland - (RS,PW,RK)-UMO-2017/27/B/ST7/00885, (AP)-UMO-2019/35/D/ST7/03781.

\bmsection{Data Availability Statement}
Source data and source code will be provided by the authors at a reasonable request.

\bmsection{Disclosures}
The authors declare no conflicts of interest. 
\end{backmatter}



\end{document}